\documentclass[twocolumn]{aastex63}

\definecolor{purple}{rgb}{0.5,0,0.87}
\definecolor{teal}{rgb}{0,0.67,0.67}
\hypersetup{linkcolor=magenta,citecolor=teal,filecolor=cyan,urlcolor=teal}

\newcommand{\dft}{NGC1052-DF2}
\newcommand{\dff}{NGC1052-DF4}
\newcommand{\hipercam}{HiPERCAM}
\newcommand{\sextractor}{\texttt{SExtractor}}
\newcommand{\scamp}{\texttt{SCAMP}}
\newcommand{\swarp}{\texttt{SWarp}}
\newcommand{\noisechisel}{\texttt{NoiseChisel}}
\defcitealias{vD2019}{vD19}

\received{June 1, 2019}
\revised{January 10, 2019}
\accepted{\today}
\submitjournal{ApJ}

\shorttitle{\dff{} is undergoing tidal disruption}
\shortauthors{Montes et al.}

\begin{document}

\title{The galaxy ``missing dark matter" \dff{} is undergoing tidal disruption}

\correspondingauthor{Mireia Montes}
\email{mireia.montes.quiles@gmail.com}

\author[0000-0001-7847-0393]{Mireia Montes}
\affiliation{School of Physics, University of New South Wales, Sydney, NSW 2052, Australia}

\author[0000-0002-6220-7133]{Ra\'ul Infante-Sainz}
\affiliation{Instituto de Astrof\'{\i}sica de Canarias, c/ V\'{\i}a L\'actea s/n, E38205 - La Laguna, Tenerife, Spain}
\affiliation{ Departamento de Astrof\'isica, Universidad de La Laguna, E-38205 - La Laguna, Tenerife, Spain}

\author[0000-0002-9510-0893]{Alberto Madrigal-Aguado}
\affiliation{Instituto de Astrof\'{\i}sica de Canarias, c/ V\'{\i}a L\'actea s/n, E38205 - La Laguna, Tenerife, Spain}
\affiliation{ Departamento de Astrof\'isica, Universidad de La Laguna, E-38205 - La Laguna, Tenerife, Spain}

\author[0000-0002-3849-3467]{Javier Rom\'an}
\affiliation{Instituto de Astrof\'isica de Andaluc\'ia (CSIC), Glorieta de la Astronom\'ia, 18008 Granada, Spain}

\author[0000-0001-5292-6380]{Matteo Monelli}
\affiliation{Instituto de Astrof\'{\i}sica de Canarias, c/ V\'{\i}a L\'actea s/n, E38205 - La Laguna, Tenerife, Spain}
\affiliation{ Departamento de Astrof\'isica, Universidad de La Laguna, E-38205 - La Laguna, Tenerife, Spain}

\author[0000-0003-3249-4431]{Alejandro S. Borlaff}
\affiliation{NASA Ames Research Center, Moffett Field, CA 94035, USA}
     
\author[0000-0001-8647-2874]{Ignacio Trujillo}
\affiliation{Instituto de Astrof\'{\i}sica de Canarias, c/ V\'{\i}a L\'actea s/n, E38205 - La Laguna, Tenerife, Spain}
\affiliation{ Departamento de Astrof\'isica, Universidad de La Laguna, E-38205 - La Laguna, Tenerife, Spain}   




\begin{abstract}
The existence of long-lived galaxies lacking dark matter represents a challenge to our understanding of how galaxies form. Here, we present evidence that explains the lack of dark matter in one of such galaxies: \dff. Deep optical imaging of the system has detected tidal tails in this object caused by its interaction with its neighbouring galaxy NGC1035. As stars are more centrally concentrated than the dark matter, the tidal stripping will remove a significant percentage of the dark matter before affecting the stars of the galaxy. Only $\sim7\%$ of the stellar mass of the galaxy is in the tidal tails, suggesting that the stars of \dff{} are starting only now to be affected by the interaction, while the percentage of remaining dark matter is $\lesssim1\%$. This naturally explains the low content of dark matter inferred for this galaxy and reconciles these type of galaxies with our current models of galaxy formation. 
\end{abstract}



\section{Introduction} \label{sec:intro}

Dark matter (DM) is a key constituent in current models of galaxy formation and evolution. In fact, without the presence of DM, the primordial gas would lack enough gravity pull to start collapsing and forming new galaxies. Because this DM is thought to interact mostly gravitationally, its presence can only be inferred from its effect on the visible matter \citep[e.g.,][]{Rubin1970, Freeman1970}. 

Although transient, DM-free, stellar aggregations (also called Tidal Dwarf Galaxies; TDGs) have been known for some time \citep[see e.g.,][]{Duc2000, Lelli2015, Ploeckinger2015, Ploeckinger2018}, the existence of long-lived non-isolated galaxies lacking DM would be a challenge to the current galaxy formation paradigm. For this reason, the recent report of two old potentially stable galaxies with low (or null) DM content: [KKS2000]04 (also known as \dft) and \dff{} \citep{vD_df2, vD2019}, has triggered an intense debate about the nature of these objects. While in the case of [KKS2000]04 (\dft) its unusual properties can be simply resolved if the object is located at 13 Mpc \citep[instead of the originally assumed 20 Mpc, see e.g.,][]{Trujillo2019}, the low velocity dispersion of the globular clusters (GCs) of \dff{} is compatible with the absence of DM, even at a closer distance \citep[13.5 Mpc,][]{Monelli2019}.

An appealing alternative to explain the low DM content of \dff{} is that this galaxy is undergoing tidal stripping \citep[e.g., ][]{Ogiya2018, Nusser2020, Yang2020, Jackson2020, Maccio2020}. In fact, \dff{} is very close both in velocity ($\Delta v\sim180$ km s$^{-1}$) and in spatial projection ($\Delta R\sim 4\arcmin$) to the galaxy NGC1035. If \dff{} is undergoing tidal disruption, this could explain the apparent lack of DM as this is preferentially stripped over the stars of the galaxy, or its GC system, due to its more extended spatial distribution \citep{Smith2013, Smith2016}.

In this paper, we explore the tidal stripping scenario and whether it can explain the ``lack'' of DM observed in \dff. We explore this scenario in two ways. First, we explore the spatial distribution of the GCs in \dff. If \dff{} is being tidally stripped by NGC1035, their stripped GCs would be placed along the orbit traced by the satellite. Therefore, exploring how the spatial distribution of GCs in \dff{} is can give us clues about the present state of the galaxy. 

Second, we obtained very deep imaging of \dff{} in order to search for any signs of tidal interactions in the stellar light around this galaxy \citep[see also][]{Muller2019}. Tidal effects in dwarf galaxies have been studied in simulations \citep[e.g., ][]{Johnston1996, Read2006, Penarrubia2008, Penarrubia2009} and they have found that if, for example, tidal tails are present in the outer parts of this galaxy, their shape and direction can provide information of the current orbit of the satellite \citep[e.g.,][]{Johnston2002, Klimentowski2009}.

In this paper, we will demonstrate that both the GCs and the stellar distribution of \dff{} are in agreement with the tidal stripping scenario. We will also discuss how this stripping scenario explains the apparent ``lack'' of DM, making this galaxy perfectly compatible with the current galaxy formation paradigm.
All magnitudes in this work are given in the AB magnitude system.

\section{Data}

The data used in this paper come from three different facilities: the Hubble Space Telescope (\emph{HST}) and two ground based telescopes: the Gran Telescopio Canarias and the IAC80 telescopes. We describe the details of each observation on what follows.

\subsection{Hubble Space Telescope imaging}

\dff{} was observed with the \emph{Hubble Space Telescope} (\emph{HST}) ACS Wide-Field Channel (WFC) as part of the programmes GO-14644 and GO-15695 (PI: van Dokkum). We  retrieved the data from the MAST archive\footnote{\url{https://mast.stsci.edu/portal/Mashup/Clients/Mast/Portal.html}}. The observations consist of a total of four orbits in the F606W and eight in the F814W band. The total exposure time is $8\,240s$ for the F606W and $16\,760s$ in F814W, respectively.

To obtain the final co-added images, we follow the approach detailed below. The individual frames were pre-calibrated in a standard way, i.e. bias and dark current-subtracted, flat-fielded, and charge transfer efficiency (CTE) -corrected producing \texttt{flc} files. A preliminary luminance co-add using the images of both filters was made using \texttt{Astrodrizzle} \citep{Gonzaga2012} with a median constant sky subtraction. We used this first co-add to create a deep master mask using \noisechisel{} \citep{Akhlaghi2015, Akhlaghi2019}, later applied to each individual frame in order to derive a more robust sky background. This deep master mask was manually expanded to include any remaining light from \dff{} and other large sources in the field-of-view, and then drizzled to the field-of-view of each individual exposure. For each one of them, we subtract a robust median sky background level (using bootstrapping to avoid the effect of outliers and cosmic-rays) from masked frames \citep[see][for futher details]{Borlaff2019}. 

At this point, the rough astrometry provided by the headers of the \texttt{flc} files was refined using the \texttt{TweakReg} utility. After this, we derived the final co-add for the images with \texttt{Astrodrizzle} into astrometrically-corrected drizzled images with the F814W and F606W filters. The relevant drizzling parameters used are: \texttt{pixfrac = 1}, \texttt{kernel = lanzcos3} and \texttt{combine type = iminmed}. Note that the \texttt{combine type} option, \texttt{iminmed}, is generally the same as the median, except when the median is significantly higher than the minimum good pixel value. In those cases, \texttt{iminmed} will choose the minimum value. The final co-add of the images still shows a significant gradient across the detector, consistent in both filters, lower close to the gap between the CCDs. To mitigate this problem, we used \noisechisel{} to generate a final sky background model for the co-added images, following the optimized methodology for the detection of the low surface brightness wings of extended sources\footnote{{\textsf{Gnuastro 2.3}} Tutorial - Detecting large extended targets: https://www.gnu.org/s/gnuastro/manual/html\_node/\\Detecting-large-extended-targets.html}. The optimal configuration was set at \texttt{tilesize = 300,300}, \texttt{interpnumngb = 3}. As a final step, after this background model has been subtracted, we readjust the median sky background level by a constant once again using the deep master mask. 

We take advantage of these high-resolution \emph{HST/ACS} observations, together with the photometry from the GTC \hipercam{} data (see Sec. \ref{sec:hipercam}) to characterize the GC population of NGC1052-DF4.

\subsection{Deep ground-based imaging}
The study of low surface brightness light in deep images is extremely challenging. It is not enough to have very deep data but a careful data reduction is crucial. In fact, it is common to find ``dips'' around the brightest extended objects in very deep surveys \citep[see e.g., CFHTLS and the HSC-SSP DR1;][]{Goranova2009, Aihara2018} due to aggressive sky subtraction. In addition, the study of diffuse and extended objects is susceptible to biases due to flat-field inaccuracies and the scattered light from bright stars, to name a few.

\dff{} was observed with the Gran Telescopio Canarias (GTC) and the IAC80 telescope. Most of the reduction steps are common for both datasets and, therefore, described in the following. Specifics for each of the telescopes are discussed later. The aim of this reduction process is to correct all systematics effects introduced by the instrument/telescope as well as to calibrate the data. The whole reduction process was done within a controlled and enclosed environment as described in \citet{Akhlaghi2020}.

Deep observations require an observing strategy that produces a background as flat as possible around our galaxy target. Another key step is the dithering pattern. For this work, we followed the dithering strategy in \citet{Trujillo2016}. In short, this consists in a dithering pattern with large steps (typically the size of the source under investigation) and whenever possible rotation of the camera. This ensures we achieve the flat background required for our goals. 

Bias frames were taken the same night of the observations as part of the different observing programs. The master bias frames were created as the sigma-clipped ($3\sigma$) mean and subtracted from the original images. 

Dome flats are not suitable for our goals due to inhomogeneities in the dome illumination that can result in gradients across the image. Consequently, the flat-field frames derived here are based on the same science exposures obtained for this work. To solve this, the flat-field correction was performed in two steps. As the bias-corrected CCD frames present steep gradients\footnote{This is due to illumination inhomogeneities as well as fringing, particularly for the GTC \hipercam{} $z$ band.}, we first stack each bias-corrected image after normalizing them by its $3\sigma$-clipped mean. This produces a rough master flat-field frame which is used to correct the CCD frame. These corrected CCD images were used to build an object mask using \noisechisel's detection maps. Again, for each CCD, the masked and normalized images are combined to create the final, and more accurate, master flat-field frames for each band.

We, then, performed the astrometric calibration of the different frames. We used the \texttt{Astrometry.net} software \citep{Lang2010} to produce an approximate astrometric solution, later refining it with \texttt{scamp} \citep{Bertin2006}. Later, we performed the sky background correction by masking all signal pixels detected with \noisechisel{} for each frame. The remaining pixels are used to compute a first degree 2D polynomial fit, that is then subtracted to the entire frame. This ensures the correction of any remaining gradient while preserving local low surface brightness structures.  

Once all frames were corrected of all systematic effects and in a common astrometric solution, they were re-sampled into a common grid using \swarp{} \citep{Bertin2010} and stacked using a $3\sigma$-clipped mean into a final image, one per filter.
The photometric calibration of both sets of images, \hipercam{} and IAC80, was done using SDSS DR12 \citep{Lam2015}.

\subsubsection{GTC \hipercam{} images}\label{sec:hipercam}

Deep multiband imaging with GTC was requested through Director Discretionary Time to observe \dff{} with \hipercam. \hipercam{} \citep{Dhillon2018} is a quintuple-beam, high-speed astronomical imager able to obtain images of celestial objects in five different filters ($u$, $g$, $r$, $i$, $z$) simultaneously. The image area of each of the five CCDs is $2048\times1024$ pixels ($2.7\arcmin \times1.4\arcmin$; 1 pixel$=0.08\arcsec$) divided into four channels of $1024\times512$ pixels.
\dff{} was observed between 2019-09-04 and 2019-09-08. During these observations, there was a problem with the electronics of the $i$-band CCD making it impossible to use this band. The average seeing of the images was $1.1\arcsec$.

For \hipercam, the standard calibration of each individual frame, i.e. bias and flat-field correction, was performed in each channel of the CCD. After the flat-field correction, we assembled each set of 4 channels into a single image per filter. To do this, a fine gain correction is done by computing the average value along the adjacent pixels for the different channels.

We obtained $\sim55$ exposures per filter using a constant exposure time of 106 s per frame. The different exposures that went into the final images were visually inspected and rejected if rejecting those with low quality or strong gradients were still present. The final exposure times in the innermost R$<1\arcmin$ of the images are: 1.6, 1.7, 1.5 and 1.7h for the $u$, $g$, $r$, and $z$ bands, respectively. 

Unfortunately, the modest field-of-view of \hipercam{} is similar to the size of the galaxy (with a diameter of $\sim1.5\arcmin$). This precludes a reliable sky subtraction and, therefore, the analysis of the extended low surface brightness emission around \dff. Consequently, even though the GTC data is very deep, we only used them to study the GC system around \dff{} (which being point-like sources are not significantly affected by the sky subtraction). The point-like source limiting magnitudes (5$\sigma$ in $2\arcsec$ diameter apertures) are: $m_u = 27.11\pm0.06$, $m_g = 26.98\pm0.07$, $m_r =26.44\pm0.07 $ and $m_z =24.81\pm0.10 $ mag.

\subsubsection{IAC80}

In order to explore the distribution of light around \dff{} and NGC1035, we use the IAC80 telecope. The IAC80 telescope is an 82-cm telescope located at the Teide Observatory in Tenerife. The camera currently installed at its Cassegrain primary focus, and used here, is CAMELOT2 (the Spanish acronym of ``Teide Observatory Light Improved Camera''). It contains a E2V 2048$\times$2048 back illuminated chip with a pixel scale of 0.336\arcsec per pixel corresponding to a $10\farcm4\times10\farcm4$ field-of-view. This field-of-view is enough to get a reliable sky estimation around \dff. \dff{} was observed with CAMELOT2 between 2019-12-19 and 2020-02-20 in the $g$, $r$ and $i$ bands. The final exposure times are $22.33$, $23.33$ an $15.66$ h for the $g$, $r$, and $i$ bands respectively. The surface brightness limits of these images are: $\mu_g = 29.8$, $\mu_r = 29.3$ and $\mu_i =  28.8$ mag/arcsec$^2$ (3$\sigma$; $10\arcsec\times10\arcsec$ boxes). They were estimated measuring the root mean square (r.m.s) of the final masked images by randomly placing $3\,000$ boxes of $10\arcsec\times10\arcsec$ across the images.

\begin{figure*}
\begin{center}
\includegraphics[width=0.8\textwidth]{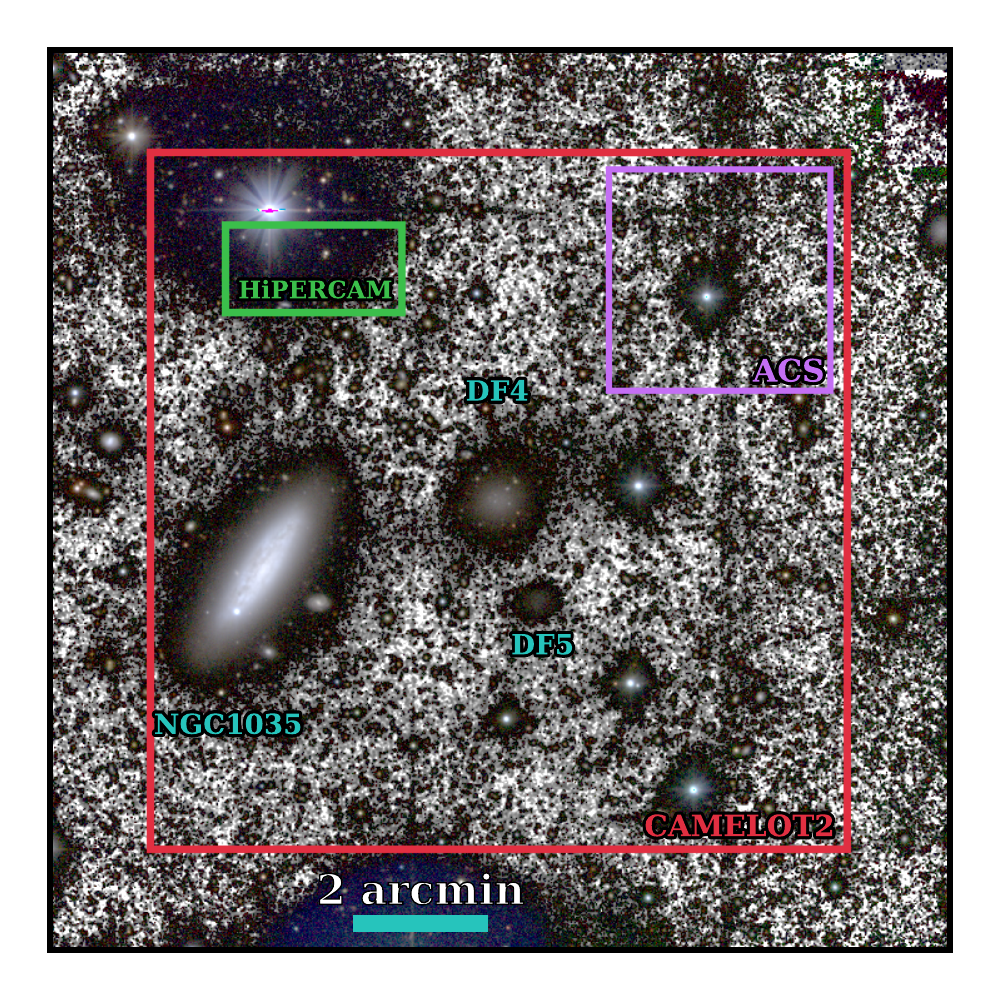}
\caption{We show an $13.3\arcmin\times13.3\arcsec$ region around \dff{} imaged with the IAC80 telescope. The figure is a composite of an RGB color image using the $g$, $r$ and $i$ IAC80 bands and a black and white $g+r+i$ image for the background. The figure highlights the main galaxies in the field-of-view: \dff, and their neighbors NGC1035 and NGC1052-DF5. The field-of-view of the different cameras used in this paper are overplotted. The surface brightness limit of this composite image is equivalent to $\mu_r\sim$29.9 mag/arcsec$^2$ ($3\sigma$; 10$\times$10 arcsec boxes). North is up, East is left. \label{fig:iac80fov}}
\end{center}
\end{figure*}

Figure \ref{fig:iac80fov} shows the $13.3\arcmin \times 13.3\arcmin$ region around \dff{} taken with the IAC80 telescope. In the figure, we have also plotted the field-of-view of the IAC80/CAMELOT2, GTC/HiPERCAM and HSC/ACS cameras for ease of comparison.

\subsubsection{PSF modelling and bright star subtraction of the IAC80 images}\label{sec:psf}

A key aspect in low surface brightness science is to remove the scattered light field produced by the brightest of objects of the images. 
To do this, an accurate characterisation of the point spread function (PSF) of the images is required. As we are only studying the distribution of light around \dff{} in the IAC80 images, this was done only for these images.

To build the extended ($R=7.5\arcmin$) PSF models of the IAC80 telescope we followed the methodology outlined in \citet{Roman2019} and \citet{Infante2020}. The outermost regions of the PSFs were obtained through dedicated observations of bright stars with magnitudes between 3.2 and 4.8 mag, in the V-band. The exposure times vary with the star magnitude and the observed band between 7 to 22s. The observations of the bright stars were conducted with a large dithering pattern in order to cover the entire field-of-view of the camera. Individual frames were masked to avoid the contamination of background sources and finally combined to create the outer parts of the PSF models. 

Because we only use bright stars to build this model, the final PSF models are saturated in the center. However, as the goal is to eliminate the contribution of the outer parts of bright stars surrounding \dff, we did not derive the inner parts of the PSF model. We fit and subtracted 8 bright stars in the field-of-view of the camera that could affect our photometry. To subtract the stars in our images, we follow the steps outlined in \citet{Roman2019}.

\section{The spatial distribution of the globular cluster system around \dff}

Globular clusters (GCs) are thought to form in the episodes of intense star formation that shaped galaxies \citep[see][and references therein]{Brodie2006}. Their typical masses ($10^{4-6}M_{\odot}$) and compact sizes \citep[half-light radii of a few parsecs,][]{Harris1996} make them easily observable in external galaxies and, therefore, good tracers of the properties of their host system even at large radial distances, unlike stellar dynamics studies that are limited to the inner regions of galaxies. 

The spatial distribution of the GC system of a galaxy might reveal clues about its present state. For instance, in an accretion event, the stripped material is placed along the orbit of the parent satellite galaxy, forming tidal streams \citep[e.g.,][]{Toomre1972, Johnston1996}. Therefore, during this process the stripped GCs will also align along the orbit \citep[e.g.,][]{Mackey2010, Hughes2019} and, consequently, their observed spatial distribution can indicate the direction of this orbit. 

\citetalias{vD2019} identified seven GCs associated with \dff{} by confirming them spectroscopically after a first selection based on their HST F606W $-$ F814W color. In this section, we use our deep \hipercam{} imaging to identify additional GC candidates and determine the spatial distribution of the GCs of \dff{}. The goal is to explore whether the GCs align through a particular direction, which could indicate that \dff{} is undergoing tidal stripping or, on the contrary, the GCs are more spherically distributed around \dff{} as it would be expected if the galaxy is in isolation.

\subsection{GCs selection}

\begin{figure*}
\begin{center}
\includegraphics[width=\textwidth]{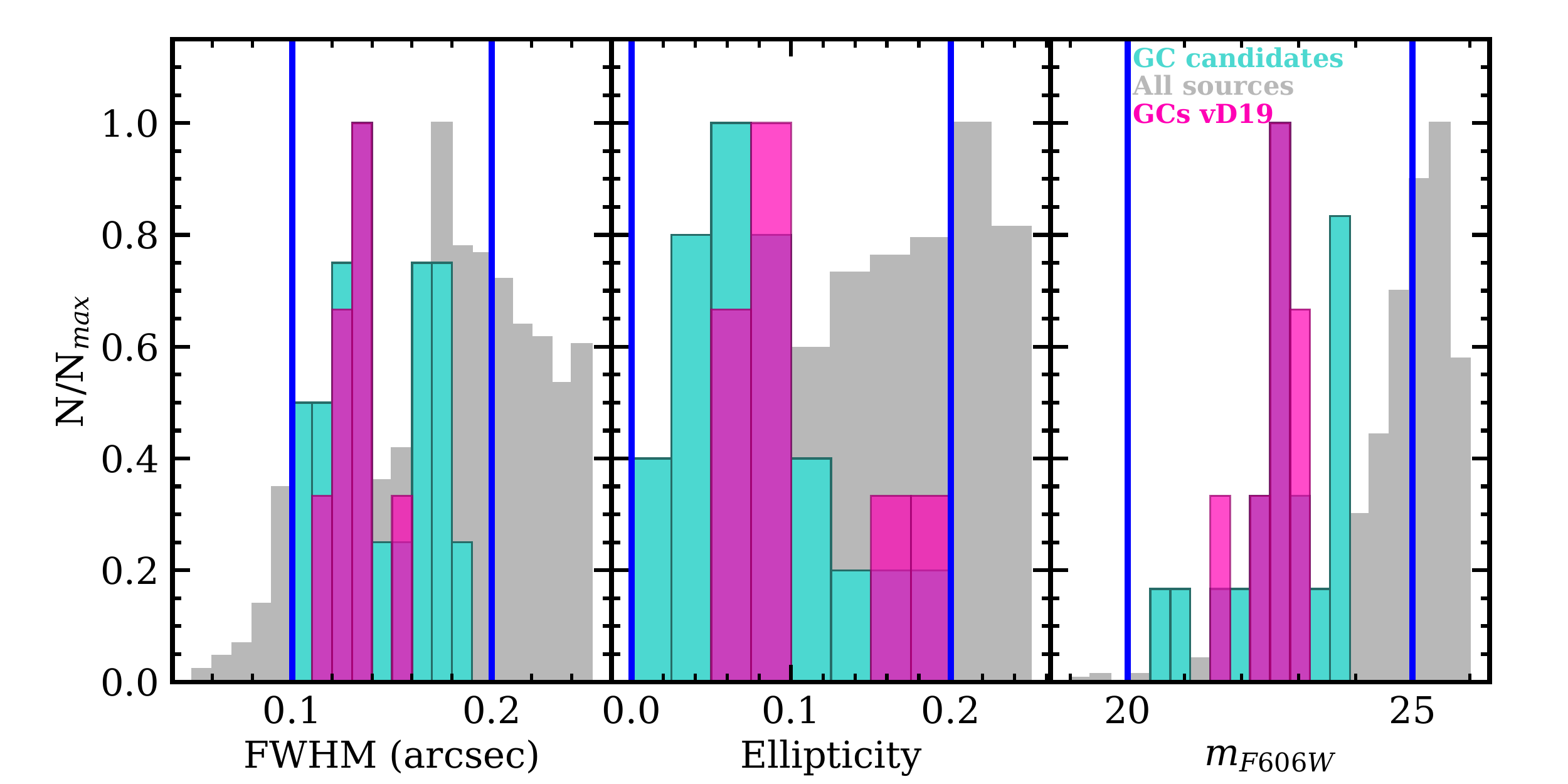}
\caption{Histograms of FWHM (left), ellipticity (middle) and F606W magnitude $m_{\rm F606W}$ (right) for all the sources simultaneously detected sources in the \emph{HST} and \hipercam{} images (grey). The magenta histograms show the distribution of the spectroscopically confirmed GCs in \citetalias{vD2019}. The blue vertical lines mark the range in each parameter of the initial selection. The mint green histograms show the sources that simultaneously satisfy the conditions of FWHM, ellipticity and magnitude enclosed by the vertical blue lines.\label{fig:hist_gcs}}
\end{center}

\end{figure*}

The task of identifying GCs in galaxies is non-trivial. The uncertain level of foreground and background contamination limits its utility to characterize the overall GC system of the host galaxy with a large degree of confidence. A common technique for identifying GCs in galaxy systems is to use their colors. However, recent works have shown the increased effectiveness of identifying GCs using more information from their spectral energy distributions \citep[e.g.,][]{Montes2014, Munoz2014}.

For this reason, in order to identify more GCs candidates in \dff{} we use the ($u$-$r$)-($r$-$z$) color-color diagram, as described in \citet{Taylor2017}. In this case, the effectiveness of this selection criteria is based in the inclusion of the $u$-band, that allows a cleaner separation of metal-poor GCs from the foreground stars of our Galaxy. The presence of hot horizontal branch stars, contributing to the flux in the $u$-band, in these metal-poor GCs will make them bluer in $u$ - $r$ than foreground stars for a given $r$-$z$.

For a preliminary selection, we took advantage of the high spatial resolution of \emph{HST} to pre-select potential GCs based on their morphology (full-width-at-half-maximum, FWHM, and ellipticity) and F606W magnitude, $m_{\rm F606W}$. This selection was modeled upon the properties of the confirmed GCs in \citetalias{vD2019}. First, we run \sextractor{} \citep{Bertin1996} on the \emph{HST} images in dual-mode using F814W as the detection image. We also ran \sextractor{} in the \hipercam{} images of \dff{} using in this case the $r$ image as our detection image. Once the catalogues for the different instruments, \emph{HST} and \hipercam, are obtained, we matched them based on their position in the sky. 

The initial GC selection consists of objects which simultaneously fulfil the following properties: $0.1\arcsec<$ FWHM $<0.2\arcsec$, $0<$ ellipticity $<0.2$ and $20<m_{\rm F606W}<25$ mag. This observed magnitude range corresponds to the absolute magnitude range $-10.65<M_{\rm F606W}<-5.65$ if the galaxy is at 13.5 Mpc or $-11.5<M_{\rm F606W}<-6.5$ if the galaxy is at 20 Mpc. The aim of this selection is to find GCs that are similar to the confirmed GCs while allowing for sources that might have been missing from the spectroscopic based sample. 
Fig. \ref{fig:hist_gcs} shows the histograms of FWHM, ellipticity and $m_{\rm F606W}$ of all the sources detected with \sextractor{} (grey), the confirmed GCs in \citetalias{vD2019} (magenta) and the GC candidates based in our preliminary selection (mint green). Twenty objects fall within this pre-selection, including the seven objects in \citetalias{vD2019}.

\begin{figure}
\begin{center}
\includegraphics[width=0.5\textwidth]{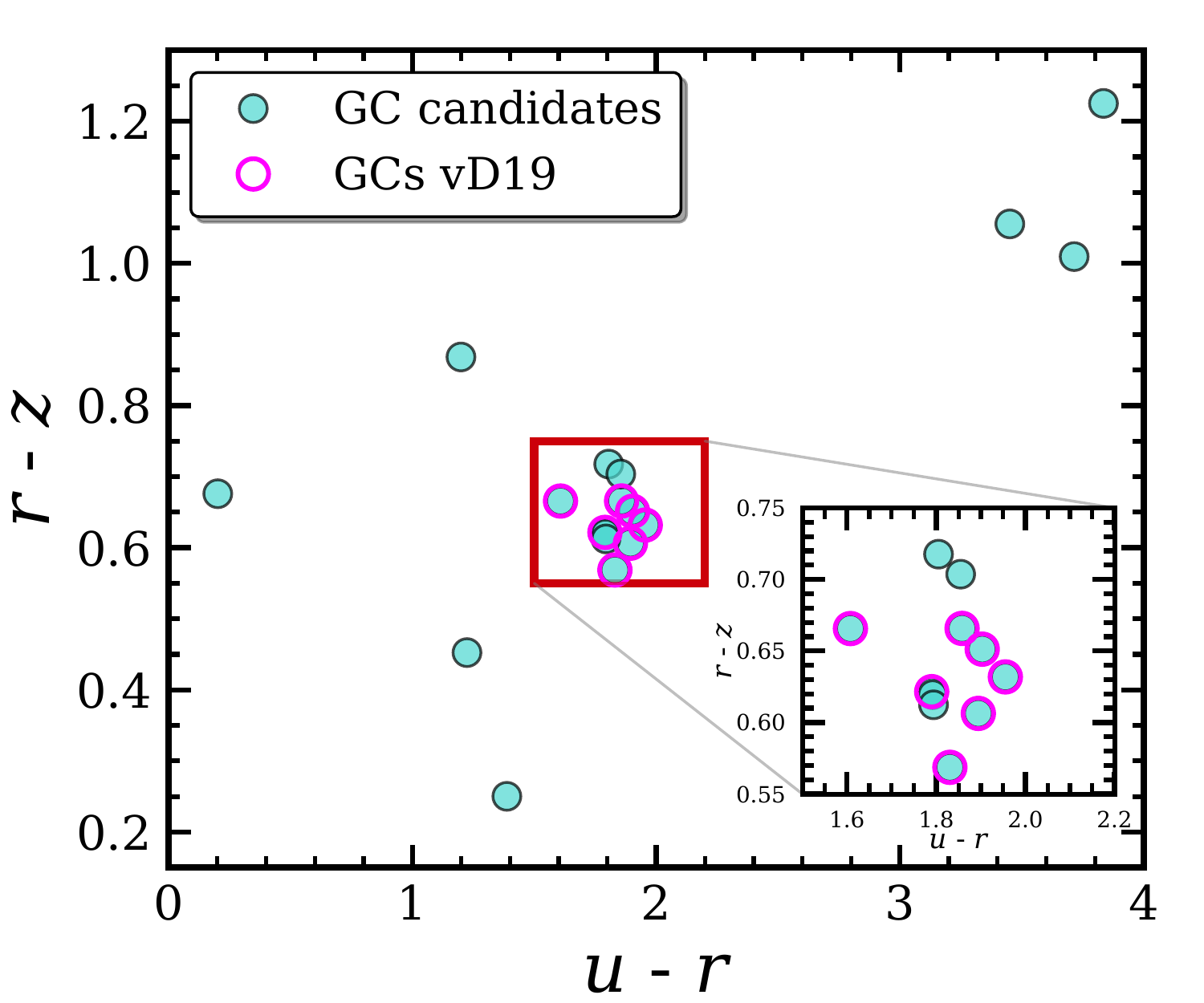}
\caption{The ($u$-$r$)-($r$-$z$) color-color diagram of the initial sample of candidate GCs (mint green). The spectroscopically confirmed GCs of \citetalias{vD2019} are highlighted in magenta. The red box indicates the color-color region we have selected to create our final sample of GC candidates. The inset shows a zoom-in into the red box for ease of viewing. \label{fig:color_color}}
\end{center}
\end{figure}

The next step is to narrow down the selection based on the colors of the GCs. Fig. \ref{fig:color_color} shows the ($u - r$)$-$($r - z$) color-color diagram for the preselected GC candidates (mint green) from our preliminary selection. We also overplotted the seven GCs in \citetalias{vD2019} as the magenta squares. As seen in Fig. \ref{fig:color_color}, the confirmed GCs fill a very narrow space in the color-color diagram and, therefore, we further narrowed down the initial selection to objects with $1.5<u-r<2.2$ and $0.55<r-z<0.75$, shown as the red box in Fig. \ref{fig:color_color}. These ranges in color are based on the $\pm3\sigma$ around the median colors of the confirmed GCs (magenta squares), with $\sigma$ being their dispersion in color.
Using this color selection, our final sample consists of a total of 11 GCs, four new candidates in addition to the confirmed GCs in \citetalias{vD2019}. The radius containing half of the GCs is $\sim155\arcsec$, $9$ times larger than the $R_e$ of the stellar light ($17\arcsec$; \citetalias{vD2019}).

The new GC candidates reported here are located to the West of \dff{}, the side facing away from NGC1035, which is less likely to be contaminated by the GC system of the disk galaxy. In addition, we expect very low GC contamination from this low-mass disk galaxy. Using the Milky Way as a reference, and the correlation between the size of the GC system and the stellar mass of the host \citep{Forbes2017}, we expect NGC1035 to have half of their GCs contained within a radius of $<4$ kpc, or $61\arcsec$/$42\arcsec$ \citep{Hudson2018} at $13.5$/$20$ Mpc, respectively. For reference, the projected distance between the centers of the galaxies is $222\arcsec$.

The coordinates and magnitudes of all the objects are listed in Table \ref{tab:1}. The magnitudes are corrected by the extinction of our Galaxy : $A_v = 0.11$, $A_g = 0.08$, $A_r = 0.06$, $A_z = 0.03$, $A_{\rm F606W} = 0.06$ and $A_{\rm F814W} = 0.04$.

 \begin{deluxetable*}{lcccccccc}[t]
 \tabcolsep=0.1cm
\tablecaption{\label{tab:1}
	Confirmed and candidate GCs of \dff.}
\tablehead{ Name & RA & DEC & $m_u$ & $m_g$ & $m_r$ &  $m_z$ & $m_{\rm F606W}$ & $m_{\rm F814W}$ \\
&  &  & $\mathrm{mag}$ & $\mathrm{mag}$ & $\mathrm{mag}$ & $\mathrm{mag}$ & $\mathrm{mag}$ & $\mathrm{mag}$ }
\startdata
C-1$^a$ & $2^h39^m14.52^s$ & $-8\arcdeg08\arcmin27.56\arcsec$ & $25.69 \pm 0.19$ & $24.35 \pm 0.03$ & $23.88 \pm 0.04$ & $23.17 \pm 0.08$ & $23.78 \pm 0.02$ & $23.46 \pm 0.02$ \\
C-2$^*$ & $2^h39^m16.96^s$ & $-8\arcdeg08\arcmin04.35\arcsec$ & $25.00 \pm 0.10$ & $23.49 \pm 0.01$ & $23.04 \pm 0.02$ & $22.41 \pm 0.04 $ & $22.93 \pm 0.01 $ & $ 22.55 \pm 0.01$ \\
C-3$^*$ & $2^h39^m18.22^s$ &$ -8\arcdeg07\arcmin23.17\arcsec$ & $24.40 \pm 0.06$ & $22.97 \pm 0.01$ & $22.50 \pm 0.01$ & $21.84 \pm 0.02 $ & $ 22.45 \pm 0.01 $ & $ 22.06 \pm 0.01$ \\
C-4$^*$ & $2^h39^m17.23^s$ & $-8\arcdeg07\arcmin05.75\arcsec$ & $24.47 \pm 0.06$ & $23.12 \pm 0.01$ & $22.68 \pm 0.01$ & $22.06 \pm 0.03 $ & $ 22.63 \pm 0.01 $ & $ 22.24 \pm 0.01$  \\
C-5$^*$ & $2^h39^m15.24^s$ & $-8\arcdeg06\arcmin57.94\arcsec$ & $23.74 \pm 0.03$ & $22.34 \pm 0.01$ & $21.91 \pm 0.01$ & $21.34 \pm 0.02 $ & $ 21.70 \pm 0.01 $ & $ 21.25 \pm 0.01$ \\
C-6$^b$ & $2^h39^m14.39^s$ & $-8\arcdeg06\arcmin59.81\arcsec$ & $23.74 \pm 0.03$ & $22.38 \pm 0.01$ & $21.95 \pm 0.01$ & $21.33 \pm 0.01 $ & $ 21.87 \pm 0.01 $ & $ 21.48 \pm 0.01$ \\
C-7$^*$ & $2^h39^m15.21^s$ & $-8\arcdeg06\arcmin52.24\arcsec$ & $24.88 \pm 0.09$ & $23.45 \pm 0.01$ & $22.99 \pm 0.02$ & $22.38 \pm 0.04 $ & $ 22.88 \pm 0.01 $ & $ 22.40 \pm 0.01$ \\
C-8 & $2^h39^m16.34^s$ & $-8\arcdeg06\arcmin44.37\arcsec$ & $25.53 \pm 0.16$ & $24.24 \pm 0.03$ & $23.73 \pm 0.03$ & $23.12 \pm 0.08 $ & $ 23.66 \pm 0.02 $ & $ 23.25 \pm0.01$ \\
C-9$^*$ & $2^h39^m12.52^s$ & $-8\arcdeg06\arcmin40.50\arcsec$ & $24.38 \pm 0.06$ & $22.99 \pm 0.01$ & $22.53 \pm 0.01$ & $21.85 \pm 0.02 $ & $ 22.46 \pm 0.01 $ & $ 22.04 \pm 0.01$  \\
C-10 & $2^h39^m10.55^s$ & $-8\arcdeg05\arcmin45.92\arcsec$ & $24.53 \pm 0.06$ & $23.23 \pm 0.01$ & $22.67 \pm 0.01$ & $21.97 \pm 0.03 $ & $ 22.67 \pm 0.01 $ & $ 22.27 \pm 0.01$ \\
C-11$^*$ & $2^h39^m16.79^s$ & $-8\arcdeg06\arcmin15.91\arcsec$ & $22.00 \pm 0.01$ & $21.12 \pm 0.01$ & $20.39 \pm 0.01$ & $19.73 \pm 0.01$ & $22.35 \pm 0.01$ & $21.92 \pm 0.01$ \\
\hline
\multicolumn{9}{c}{$^a$This candidate might be associated to NGC1052-DF5.} \\
\multicolumn{9}{c}{$^b$ This candidate has been identified as a star in \citet{Shen2020}.} \\
\multicolumn{9}{c}{$^*$GC spectroscopically confirmed in \citetalias{vD2019}.} 
\enddata
\end{deluxetable*}

\subsection{The distribution of GCs around \dff}\label{sec:axis_gcs}

\begin{figure}
 \centering
   \includegraphics[width = 0.5\textwidth]{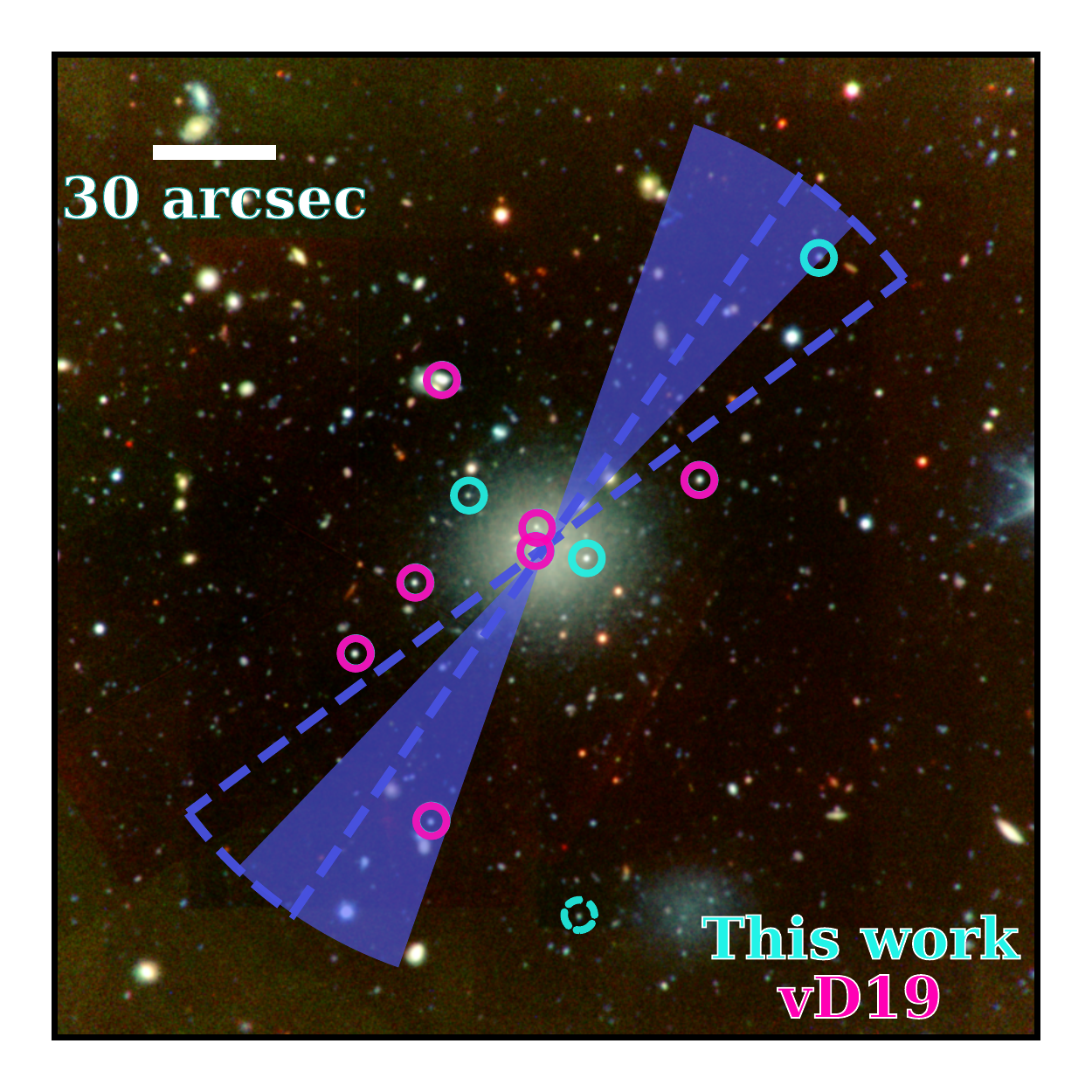}
   \caption{\hipercam{} postage stamp of $240\arcsec\times240\arcsec$ around \dff. The mint green circles highlight the candidate GCs found in this work while the magenta circles are the GCs found in \citetalias{vD2019}. The dashed mint circle at the bottom points to a GC probably associated with NGC1052-DF5. The blue area indicates the range of directions of the orbit of \dff{} defined by the spatial distribution the GCs. The area enclosed by the dashed blue line is the same but removing the GC probably associated with NGC1052-DF5 (dashed mint circle).
   In this image, $30\arcsec$ corresponds to a distance of $\sim2$ kpc at a distance of 13.5 Mpc and $2.9$ kpc at 20 Mpc.} \label{fig:gcs}
    \end{figure}

In this section, we address the spatial distribution of the GCs in \dff{} in order to obtain clues about the interaction state of the galaxy. As mentioned before, in an accretion event, the stripped material (stars and GCs) will be deposited along the orbit of the parent satellite. 
In Fig. \ref{fig:gcs}, we show an RGB image created using the \hipercam{} $g$, $r$ and $z$ bands. The highlighted sources are the confirmed GCs from \citetalias{vD2019} (magenta) and the new candidates obtained in the previous section (mint green). The GCs tend to align in a particular axis along the galaxy.

\begin{deluxetable}
{lc}[t]
 \tabcolsep=0.15cm
\tablecaption{\label{tab:2}
	Angles and their errors of the axes defined by the spatial distribution of the GCs of \dff. These angles are defined from the X-axis counter-clockwise.}
\tablehead{ & Angle $\pm$ Error }
\startdata
All the GCs in this work &  $ 58\pm12$\\
Removing DF5's GC &  $46\pm9$\\
Confirmed GCs in \citetalias{vD2019} & $50\pm18$
\enddata
\end{deluxetable}

In order to derive this axis, we use \texttt{scipy.odr} to perform a orthogonal distance regression to all the GCs, confirmed and candidates. We use the orthogonal distance regression as we are interested in calculating the \emph{orthogonal} distance to the axis instead of predicting the value of Y for any given X. The range of possible axes, i.e. directions of the orbit, will be given by the slope of the regression and its error. This is indicated by the blue shaded region in Fig. \ref{fig:gcs}. This region encompasses $68\%$ of the axes compatible with the GC distribution. One of the GCs, marked by a dashed circle, might be associated to NGC1052-DF5 given its proximity to this galaxy and, consequently, might be biasing the computed axis. Taking that into consideration, we computed again the range of possible axes but this time removing this GC, shown by the region enclosed by the dashed blue line. These two regions are compatible given the small number of sources. They are also compatible with the range of axes calculated if only using the confirmed GCs in \citetalias{vD2019}. The angles defined by these axes and their errors, derived from the slopes of the orthogonal distance regressions, can be found in Table \ref{tab:2}\footnote{Removing C-6, identified as a star in \citet{Shen2020} does not change the results.}.

Consequently, the non-isotropic distribution of GCs around \dff{} is suggesting that it might be experiencing tidal interaction with a neighboring galaxy. 

\section{Tidal stripping of the stellar distribution}

While the study of the spatial distribution of the GCs in \dff{} can provide insights on whether the galaxy is undergoing tidal stripping, the strongest evidence is contained in the outermost part of the object. In fact, in the case of a tidal stripping event, the stellar distribution of \dff{} is expected to be distributed following a characteristic S-shape \citep[e.g.,][]{Johnston2002, Klimentowski2009}. We explore this using the IAC80 ultra-deep imaging.

\subsection{Removal of NGC1035}

Fig. \ref{fig:iac80fov} already shows that the presence of an excess of light around \dff{} is evident. This excess is centered around the galaxy and has the characteristic S-shape of an object undergoing tidal stripping \citep[e.g.][]{Johnston2002}. In order to study how this light is distributed, it is crucial to remove possible contaminants that can bias our results. In Sec. \ref{sec:psf}, we have already modelled and subtracted the bright stars in the field-of-view, including the nearest star to the West of \dff. However, even though NGC1035 is $\sim4\arcmin$ away, the faint halo of this galaxy can contaminate the light of \dff{} in very deep observations.

In order to eliminate this source of contamination, we modelled the galaxy using \texttt{ellipse} and \texttt{bmodel} in IRAF. While \texttt{ellipse} fits elliptical isophotes to the image, \texttt{bmodel} uses its output to create a 2D model of the galaxy. The advantage is that the ellipticity and position angle can vary at different radius, achieving a more authentic model of the galaxy and minimizing the residuals.

Prior to using \texttt{ellipse}, we need to carefully mask foreground and background sources to reduce contamination that can affect the fitting procedure. As a single \sextractor{} setup for the detection and masking in deep observations is not appropriate, we used a two-step approach; a ``hot+cold'' mode \citep[e.g.,][]{Rix2004, MT14}. The ``cold'' mode will mask extended sources while the ``hot'' mode is optimized to detect the more compact and faint sources that are embedded within the galaxy light. 

\sextractor{} was run in a deep combined $g+r+i$ image and both masks were built from the generated segmentation maps. In the case of the ``hot'' mode, we unsharp-masked the original image in order to facilitate the masking of the dusty lanes of the galaxy. To create this unsharp-masked image, we convolved the image with a box filter with a side of 6 pixels and then subtracted it from the original. The ``cold'' mask was further expanded 6 pixels while the ``hot" was expanded 3 pixels. Both masks were combined to create the final mask for our IAC80 images, unmasking NGC1035 on the ``cold'' mask. This guarantees that the dust lanes and other sources within the galaxy are masked but not the diffuse stellar light of NGC1035. The final mask was visually inspected to manually mask any remaining light that was missed by the process described above. 

We run \texttt{ellipse} in the deep $g+r+i$ image in two steps. In the first run, all the parameters were allowed to vary freely while in the second run we fixed the centre of the galaxy to the median centers returned in the first run. The geometry of the ellipses obtained in the fitting of the deep image were then used to measure the individual $g$, $r$ and $i$ images and create a model of the galaxy using \texttt{bmodel}\footnote{The results of the fit are consistent if we model each band separately.}. The fit performed by \texttt{ellipse} reaches down to $\sim3\arcmin$ from the centre of NGC1035.

 \begin{figure*}
 \centering
   \includegraphics[width = \textwidth]{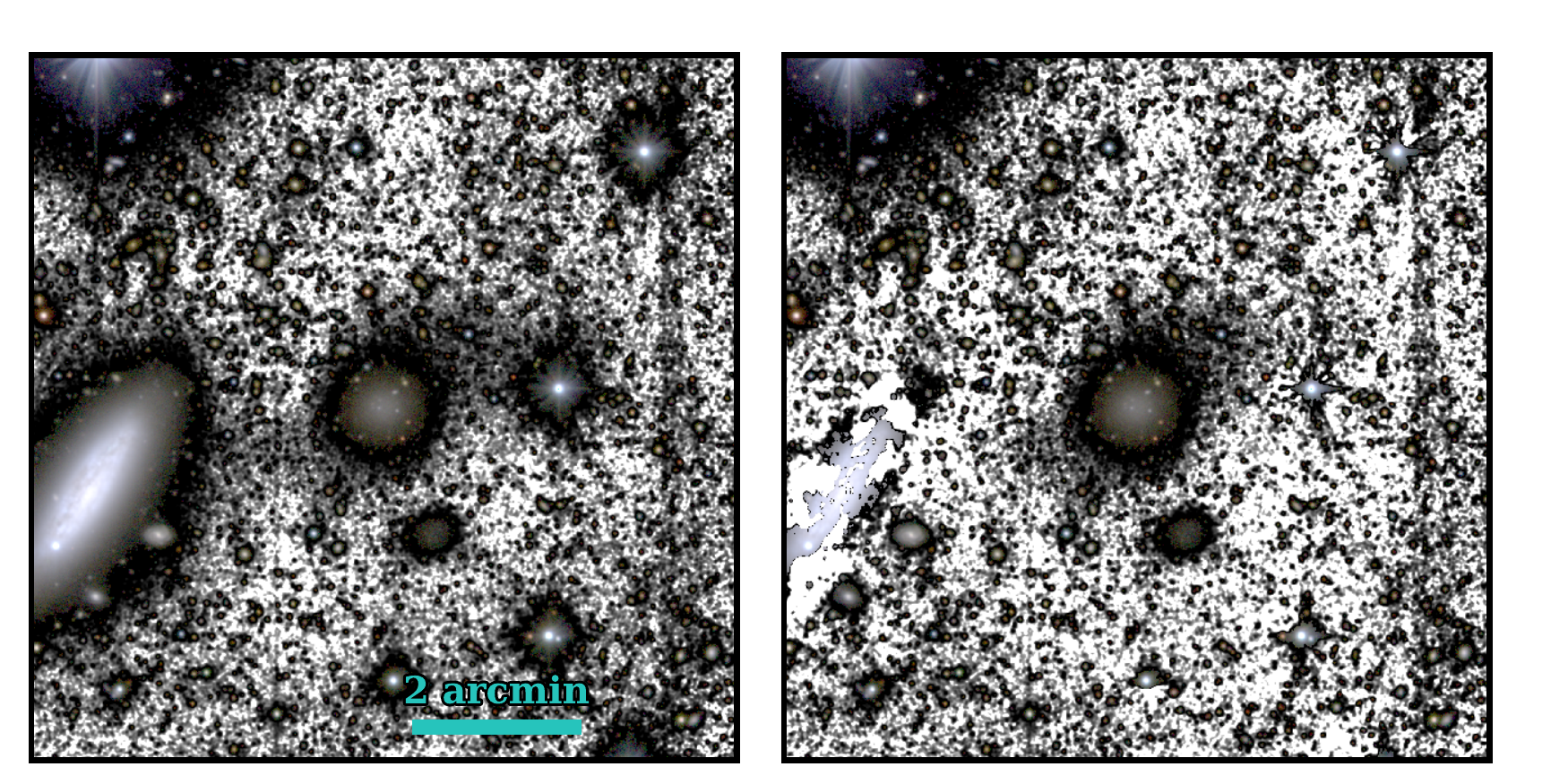}
   \caption{Left panel shows the original IAC80 postage stamp of the $500\arcsec\times500\arcsec$ around \dff. The right panel shows the results of subtracting the nearest bright star (west) and the fitting and removal of NGC1035. After subtracting these sources of contamination the S-shape of the galaxy becomes more evident. } \label{fig:removal}
    \end{figure*}

Fig. \ref{fig:removal} shows the result of subtracting both the light of NGC1035 and the nearest bright star toward the west of the image. The S-shape of the light around \dff{} becomes even more conspicuous after subtracting possible contaminants.

\subsection{Surface brightness radial profiles of \dff}\label{sec:profs}

The goal of this paper is to investigate whether \dff{} is interacting with NGC1035 and if that could explain the lower content of DM in this ultra-diffuse galaxy. The presence of tidal tails and other asymmetries that indicate interactions will appear as an excess of light at large radius and deviations from the morphology of the inner parts of the galaxy. 
To that end, we derived the radial profiles for the IAC80 images using the software \texttt{ellipse} in IRAF. \texttt{ellipse} fits elliptical isophotes to the 2-D images of galaxies using the method described in \citet{Jedrzejewski1987}. It provides the median intensity, ellipticity and position angle for each of the fitted isophotes. 

   \begin{figure*}
   \centering
   \includegraphics[width = 0.99\textwidth]{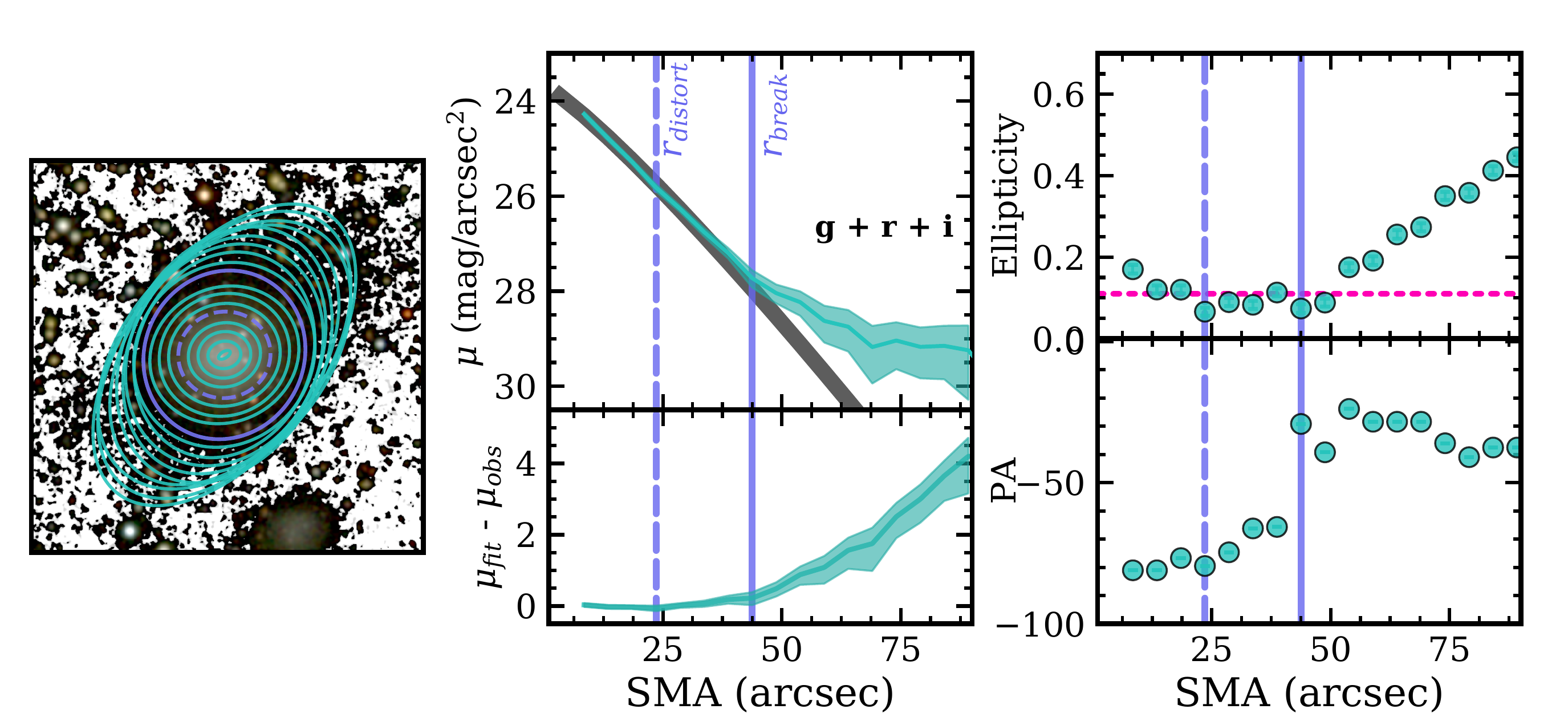}
   \caption{Output from \texttt{ellipse} for the $g+r+i$ image of \dff. The left panel shows the $200\arcsec\times200\arcsec$ region around \dff{} with the fitted ellipses overplotted. The middle panel shows the radial surface brightness profile as a function of the semi-major axis (top) and the residuals from fitting this profile with a S\'ersic model (bottom). The rightmost panel presents the profiles of ellipticity (top) and PA (bottom). }\label{fig:ellipse}%
    \end{figure*}

We used the same methodology than in the previous section to build the mask for fitting \dff. We run \sextractor{} in the deep $g+r+i$ image and obtain the segmentation maps to build the masks. In this case, the ``hot'' model was run in the original $g+r+i$ image. The ``cold'' mask was further expanded 5 pixels while the ``hot" was expanded 3 pixels, and \dff{} was left unmasked on the ``cold'' mask. That leaves the light of the galaxy unmasked while masking all compact sources like background galaxies and GCs that are superposed.

Once we have all contaminant sources masked, we derived the radial profiles for the IAC80 $g+r+i$ image with \texttt{ellipse}. This image was smoothed with a Gaussian of $\sigma = 2$ pix to improve the signal-to-noise ratio of the outer parts of the galaxy. The isophote fitting process was done in two steps: 1) allowing all parameters to vary freely and 2) fixing the centers to the median centers of the isophotes returned by the first iteration. The \texttt{ellipse} geometry obtained was then used to extract the photometry from the original $g+r+i$ image (no smoothing applied) and for each of the individual bands.

Fig. \ref{fig:ellipse} shows the output of \texttt{ellipse} for the original $g+r+i$ image. In the leftmost panel, a postage stamp of a $200\arcsec\times200\arcsec$ region around \dff{} is shown with the fitted ellipses overplotted. The 1-D radial surface brightness profile as a function of the semi-major axis (SMA) is shown in the top middle panel, up to $\sim90\arcsec$. The shaded regions represent the r.m.s error in each elliptical isophote. The right panel shows the ellipticity (top) and PA (bottom) as a function of SMA. The magenta dashed line in the ellipticity profile marks the value of the ellipticity of \dff{} as reported in \citetalias{vD2019}. The profiles for each of the individual bands are shown in Appendix \ref{app:profiles}.

\subsection{Defining $r_{break}$ and $r_{distort}$ in \dff}\label{sec:radii}

One of the most striking features in the 1-D surface brightness profile of \dff{} is its sudden change in the slope at SMA $\sim45\arcsec$. In order to investigate this, Fig. \ref{fig:ellipse} also shows a \citet{Sersic1968} fit to the 1-D surface brightness profile of \dff{} (grey line in the top middle panel). This S\'ersic model ($R_e=18\arcsec$\footnote{This is $1.2$ kpc at $13.5$ Mpc or $1.7$ kpc at $20$ Mpc.} and n$=0.86$) fits nicely the inner ($R<40\arcsec$) parts of the galaxy, as mentioned in \citetalias{vD2019}, while at large radius is not a good description of the surface brightness profile.
The bottom middle panel shows the residuals of subtracting the best S\'ersic fit from the $g+r+i$ surface brightness. At a radius of $\sim45\arcsec$ there is a significant excess of light with respect to the model fit. Although this excess starts earlier, the slope of these residuals at this radius becomes steeper. This is accompanied by a steep increase in the ellipticity of the isophotes reaching a value of $\sim0.4$ at $\sim90\arcsec$. In addition, there is a sudden change in the PA trend.

Following \citet{Johnston2002}, we tentatively define the radius where the surface brightness profile of \dff{} departs ($\Delta\mu \gtrsim 0.2$ mag/arcsec$^2$) from the S\'ersic model as the break radius ($r_{break}$) marked as the solid vertical blue line in the profiles at $43.8\arcsec$. We also marked this particular ellipse in the postage stamp (left panel Fig. \ref{fig:ellipse}).

Although this break is the most prominent feature in the surface brightness profile, the satellite starts losing its shape well within this point. \citet{Johnston2002} refers to this point where the morphology of the galaxy is affected by tidal forces as $r_{distort}$. In the case of \dff{}, at $\sim23.5\arcsec$ the PA starts increasing, i.e. twisting of the isophotes, up to $r_{break}$. This radius is where we define $r_{distort}$ and it is marked in Fig. \ref{fig:ellipse} as the dashed blue vertical line in the profiles and the dashed blue ellipse in the postage stamp.

\subsection{The S-shape of \dff}

\begin{figure*}
   \centering
   \includegraphics[width = \textwidth]{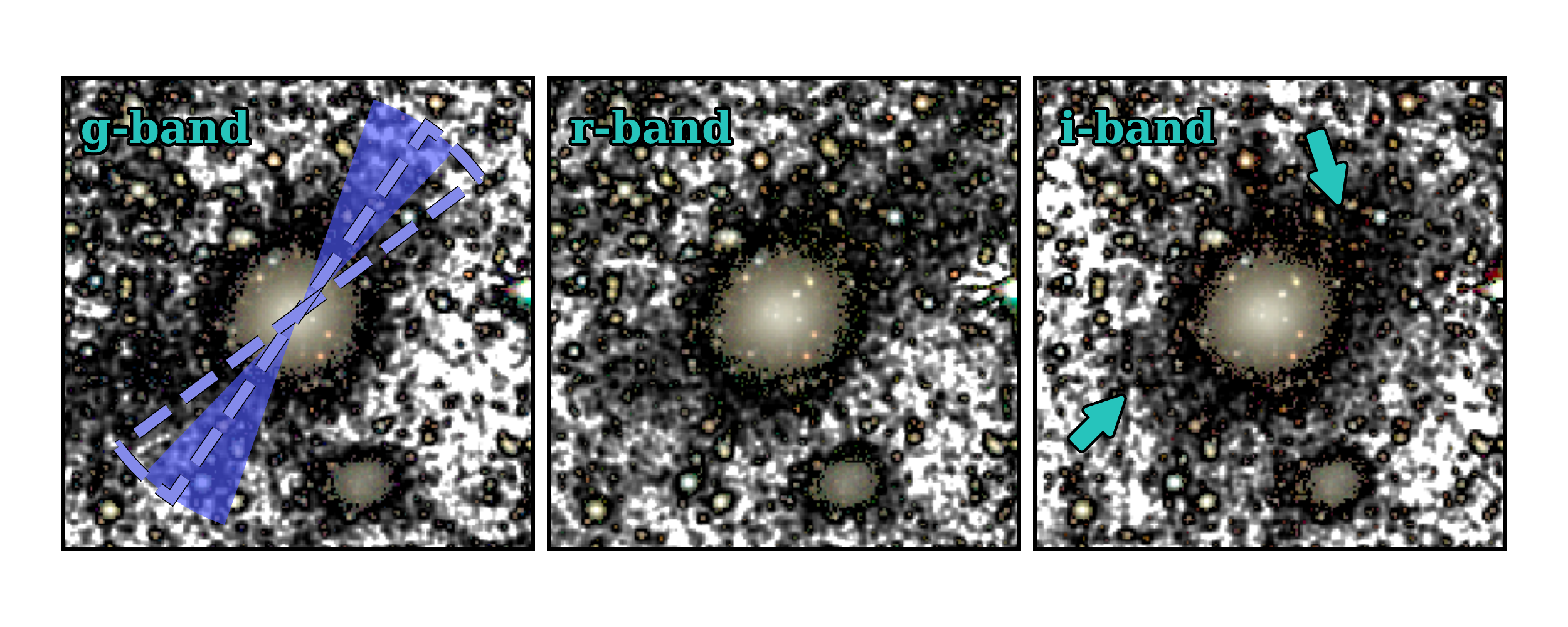}
   \caption{Postage stamps of the $250\arcsec\times250\arcsec$ region centered in \dff{}, for the $g$ (left), $r$ (middle) and $i$ (right) IAC80 images. The blue-shaded region indicates the range of the different directions of the orbit of this galaxy as derived by the spatial distribution of GCs in Sec. \ref{sec:axis_gcs}, over-plotted on the $g$-band stamp. The region outlined by the dashed line corresponds to the same range of orbits but without including the GC possibly associated with NGC1052-DF5. The excess of light to the North-West and South-East of \dff{} is marked by two teal arrows in the rightmost panel ($i$-band). } \label{fig:stamps}%
    \end{figure*}

In simulations, tidal debris is seen to spread along the orbit of the satellite in thin streams \citep[e.g.,][]{Johnston1996}. This suggests that the morphology of these loose populations can be used to constrain the direction in which the satellite is moving, which in turn can tell us something about the satellite’s orbit.
In the previous section, we have seen that the change of shape of \dff{} with radius in the $g+r+i$ image suggests that the outer parts of this galaxy are fully compatible with being tidally disrupted (Sec. \ref{sec:radii}). In order to assess if signs of this interaction can be seen in the individual images, we binned $4pix\times4pix$ the IAC80 images to improve the signal-to-noise ratio of the faint outer parts of the galaxy. Consequently, the new pixel is the sum of sixteen original pixels. In addition, we convolved the images with a Gaussian kernel of $\sigma=1$ (rebinned) pix to further enhance structures. Fig. \ref{fig:stamps} shows the postage stamps of the $250\arcsec\times250\arcsec$ region centered in \dff{} of the three bands: $g$ (left), $r$ (middle) and $i$ (right). 

In each of the images, there is an excess of light around \dff{} aligned with the range of orbit direction derived from the distribution of GCs in Sec. \ref{sec:axis_gcs}, overplotted on the $g$-band stamp (left). That is, to the North-West and South-East of the galaxy, marked by the teal arrows in the rightmost panel of Fig. \ref{fig:stamps}. This excess is seen in all three bands independently. By confirming the presence of this excess in the three different bands, we ensure that it is a real feature.

To explore in more detail this observed S-shape, and therefore constrain the direction in which \dff{} is moving, we derived the isocontours of light of this galaxy. To do that, we use the rebinned IAC80 $g+r+i$ and convolve it with a Gaussian of $\sigma = 2.5$ pix to further enhance structures that might be buried in the noise. We constructed a new mask for this image by visually masking all the sources. In addition, we generously masked the residuals of the star to the west and of NGC1035. 

\begin{figure*}
   \centering
   \includegraphics[width = \textwidth]{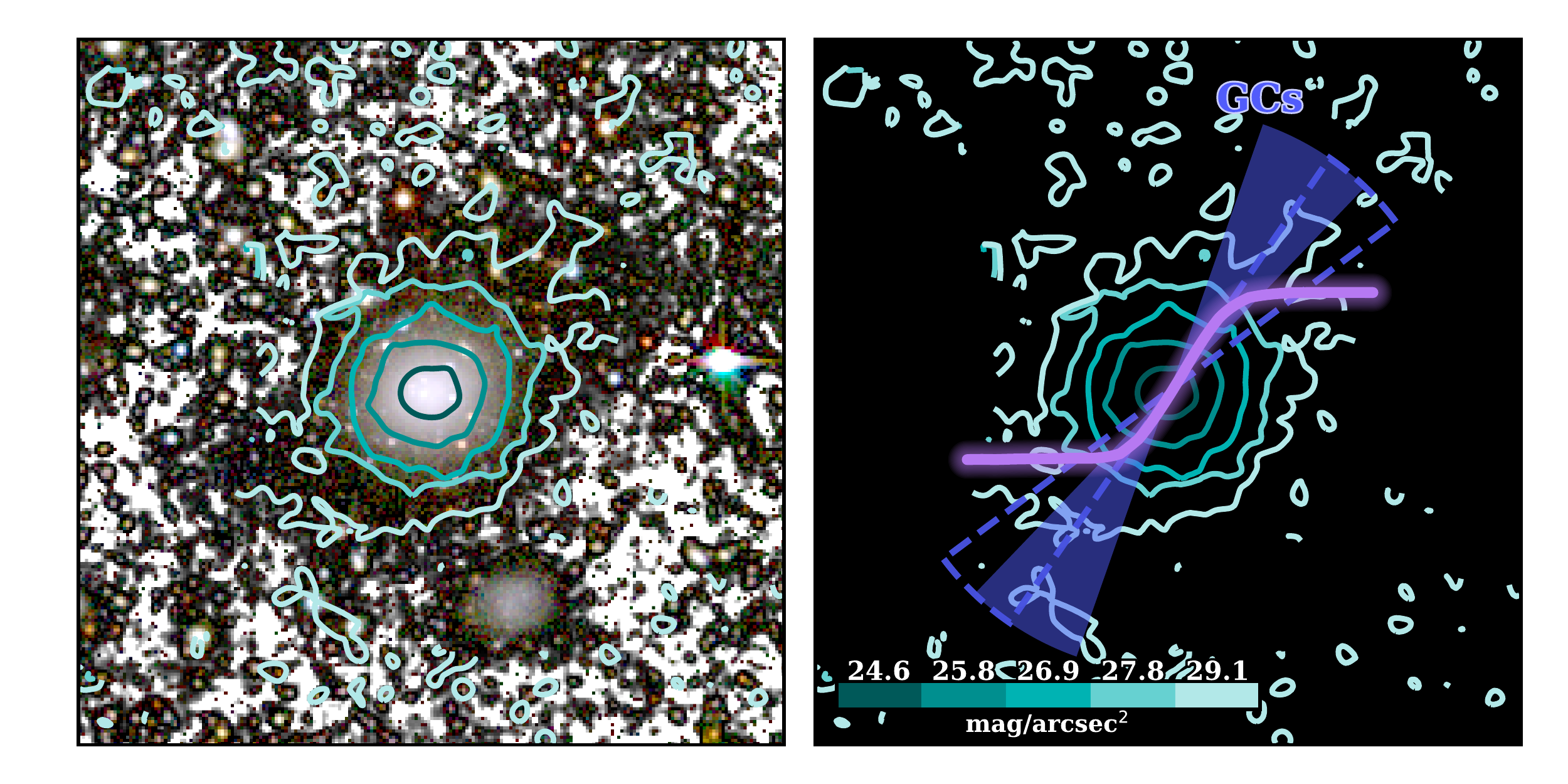}
   \caption{Contours at $5$ different surface brightness levels of \dff. The left panel shows the contours overplotted on a composite from an RGB color image and a black and white $g+r+i$ image for the background. The images are $4\times4$ binned. The right panel shows the contours over a black background with a tentative S-shape (purple solid line) tracing the shape of the last contours (at $29.1$ mag/arcsec$^2$). We also plotted the range of directions of the orbit of this galaxy obtained from the GC distribution (blue solid area) and without the GC probably associated with NGC1052-DF5 (dashed line) in Sec. \ref{sec:axis_gcs}. The S-shape is compatible with \dff{} moving along the direction defined by the position of the GCs.} \label{fig:contours}%
    \end{figure*}

We derived the isocontours of light using \texttt{matplotlib}'s \texttt{contour} function in a similar way than in \citet{MT19}. \texttt{contour} provides and draws isocontour lines at different given intensities in the image. In this case, we derived the radial light profiles of the binned $g+r+i$ image in elliptical apertures, assuming a fixed ellipticity of 0.11 (b/a = 0.89) and PA of $7\deg$ (counterclockwise from the X axis), the properties of the inner parts of the galaxy\footnote{As shown in Figure 1 of \citet{MT19}, the purpose of this is to obtain approximate values at a certain distance in order to derive a contour, and therefore, the real shape of the galaxy light at that distance.}. Using this radial profile, we interpolated the intensity values at five physical radial distances: 10, 20, 30, 40 and $50\arcsec$ from the centre of \dff. The different distances correspond to surface brightness values of: 24.6, 25.8, 26.9, 27.8 and 29.1 mag/arcsec$^2$. These contours are plotted on Fig. \ref{fig:contours}. On the left panel, the underlying image is a composite of an RGB color image created using the binned ($4\times4$) $g$, $r$ and $i$ filters and a black and white $g+r+i$ binned image for the background. 

On the right panel, we plotted the same contours on a black background to facilitate visualization. We can clearly see that the last contour (lightest teal shade, at $29.1$ mag/arcsec$^2$) shows the presence of tidal tails from material moving away from \dff. Tentatively, we have drawn a schematic S-shape (in purple) tracing the shape of this last contour. We also plotted the region containing the axes from the distribution of GCs (Sec. \ref{sec:axis_gcs}). The S-shape observed here is common in simulations where a dwarf galaxy is interacting with a more massive galaxy (e.g., Figure 5 in \citealt{Johnston2002}) and also in observations \citep[e.g.,][]{Koch2012}. 

\subsection{Tidal radius}

The tidal radius identifies the radius where the tidal effects are expected to be important in the satellite. To compute the tidal radius ($r_{tidal}$) for \dff{} we follow Equation 5 in \citet{Johnston2002}. This equation is for the tidal radius at the pericenter of the satellite orbit. Instead, as we do not know where \dff{} is in relation to its orbit around NGC1035, we computed the instantaneous $r_{tidal}$ which is the tidal radius at the current position of the satellite from the parent galaxy and assuming zero ellipticity, i.e., a circular orbit \citep{Johnston2002}. Consequently, the equation for the instantaneous $r_{tidal}$ is given by

\begin{equation}
    r_{tidal,inst} = D \times \left( \frac{m_{dyn}}{M_{1035,dyn}} \right)^{1/3}
\end{equation}

where $D$ is the radial distance between the galaxies and $m_{dyn}$ and $M_{1035,dyn}$ are the dynamical masses for \dff{} and NGC1035, repectively.
The dynamical mass of NGC1035 is $M_{1035,dyn} = 1.9\pm0.1\times10^{10}M_{\odot}$ \citep{Truong2017}\footnote{The dynamical mass of NGC1035 in \citet{Truong2017} was estimated at a distance of 17 Mpc. As the way this dynamical mass was estimated is proportional to the size of NGC1035, we converted to the according distance in our calculations.} while for \dff{} we took the estimation of the total mass from \citetalias{vD2019}; $m_{dyn} = 4^{+12}_{-3}\times10^{7} M_{\odot}$, assuming a distance of 20 Mpc \citep[see also][]{Danieli2020}. At a distance of 13.5 Mpc, this dynamical mass would be around half of that value. Assuming that the projected distance between the galaxies is the actual distance between their centers, $D = 222\arcsec$, we obtained $r_{tidal} =33\pm8 \arcsec$  and $ 28\pm 12\arcsec$, depending on the distance assumed, 13.5 and 20 Mpc, respectively. 

This is similar to the $r_{distort}$ identified in Fig. \ref{fig:ellipse} ($23.5\arcsec$). $r_{distort}$ is defined as the radius where the morphology of the galaxy departs from its original shape. In this case, at $r_{distort}$ the PA of the isophotes starts increasing, i.e. twisting, marking the point where the tidal forces are starting to affect the morphology of galaxy although the stars still remain bound to the satellite \citep[e.g.,][]{Penarrubia2009}.

Note that if \dff{} was located at 20 Mpc \citep{Danieli2020}, the most likely disturber would be NGC1052. Assuming the following dynamical mass for NGC1052 $M_{1052,dyn} = 1.7\pm0.9\times10^{12}M_{\odot}$ \citep[at 18 Mpc;][]{Pierce2005} and a distance of $28\farcm7$ between both galaxies, we obtain a tidal radius of $51\pm15\arcsec$. This is around twice as large as if we assume NGC1035 is the main responsible of the tidal distortion of \dff. This larger value for the tidal radius would be in tension with the $r_{distort}$ measured in this paper. This suggests that NGC1035 is the likely responsible of the disruption of \dff{}.

\subsection{Stellar mass in the tidal tails}\label{sec:mass_tails}

$r_{break}$ identifies the radius where observations become dominated by the unbound populations of stars, and therefore where we can find breaks in the light profiles \citep[e.g.,][]{Johnston2002, Penarrubia2009}. The lack of strong distortions in the stellar light at the centre of \dff{} indicates that the stars of this galaxy might only be starting to be stripped now. 
Consequently, we can infer a lower limit to the amount of stellar mass stripped so far by \dff{} by measuring the stellar mass outside $r_{break}$. 
The fraction of the stellar light of \dff{} beyond $r_{break}$, including the tidal tails down to $\mu_r = 29.1$ mag/arcsec$^2$, is 7$\pm1\%$, averaging the individual estimates from the $g$, $r$ and $i$ bands. This relatively low value for the stripped stellar mass, together with the fact that the central part of the galaxy appears undisturbed, is compatible with the idea that the stellar body of the galaxy is only starting to be disrupted now. Note that this quantity is a lower limit of the true fraction of mass lost by the galaxy.

\section{Discussion}

In this work, we show that \dff{} presents clear signs of interaction. In the following we discuss how this interaction can affect the amount of DM present in this galaxy. For all the calculations we assume that the distance to the galaxy is $13.5$ Mpc \citep{Monelli2019}, although qualitatively the main results presented here do not depend on this assumption. 

Note that the previous deep imaging of \dff{} reached around $\mu_r = 28.5$ mag/arcsec$^2$ \citep[$3\sigma$, $10\arcsec\times10\arcsec$, ][]{Muller2019}, still inconclusive for addressing the problem of the interaction between these two systems. Similarly, the Dragonfly data where this galaxy was identified \citep{Cohen2018} reaches between $\mu_{g/r} = 27.4$ and $28.0$ mag/arcsec$^2$ ($3\sigma$, $12\arcsec\times12\arcsec$) as reported in \citet{Merritt2016}.

\subsection{The distribution of light and GCs of \dff}

Matter stripped from  satellite galaxies forms tidal tails which will eventually become completely unbound and form larger tidal extensions tracing approximately the original orbit of the satellite in its host halo. In the same manner, the GCs of these satellites will be stripped from the galaxy and deposited along the orbit.

Using N-body simulations, \citet{Klimentowski2009} studied the properties of tidal tails of a dwarf galaxy orbiting in a Milky Way-like potential. They showed that there are \emph{always} two tidal tails emanating from the two opposite sides of the dwarf galaxy and that, for most of the time, these tails are oriented radially towards the host galaxy, and not along the orbit. This shape, reminiscent of an S, is produced by the particles that once unbound are seen to move in the direction of the tidal forces, that is perpendicular to the orbit of the satellite, before being dispersed along it.

The outskirts of \dff{} present an excess of light to the North-West and South-East of the  galaxy (marked by arrows in Fig. \ref{fig:stamps}). In Fig. \ref{fig:contours}, we explored this in more \emph{depth} using a binned $g+r+i$ and plotting isocontours of light at different surface brightness values. The faintest surface brightness ($29.1$ mag/arcsec$^2$) clearly shows an S-shape (tentatively drawn in purple). The tails of the S are oriented directly to NGC1035 (see Fig. \ref{fig:iac80fov}) indicating that this disk galaxy \citep{Truong2017} is causing the tidal perturbation. As shown in Figure 5 in \citet{Monelli2019}, NGC1035 is the closest galaxy to \dff{} at a projected distance of $\sim 222\arcsec = 14.5$ kpc, at a distance of $13.5$ Mpc \citep{Sorce2014, Monelli2019}.

In addition, in Sec. \ref{sec:axis_gcs}, we identify four new GC candidates of \dff. These, along with the confirmed GCs in \citetalias{vD2019} align in a particular direction. This preferential alignment of the GCs suggests that they are also being stripped in the interaction and therefore they are distributing in the direction of the orbit.
 
Putting all the evidence together, what Fig. \ref{fig:contours} is telling us is that \dff{} is moving roughly along the direction defined by the axis of the GCs, roughly perpendicular to the tails of the S. 

\subsection{Can tidal stripping explain the missing dark matter in \dff?}

Deep imaging of \dff{} has revealed that this galaxy is undergoing tidal disruption. Now, the question that arises is whether this tidal interaction could explain the low DM content measured in this galaxy. 

\citet{Smith2013, Smith2016} used numerical simulations to explore the effect of tidal stripping in the DM content of dwarf galaxies in clusters of galaxies. They found that only when the remaining DM fraction falls under $10-15\%$ is when the stars and GCs of the galaxy are significantly stripped. We warn the reader that these simulations cannot be directly compared to the \dff{} scenario but they provide a qualitative understanding of the problem.

To assess whether the tidal stripping scenario explains the properties of \dff, we need to calculate the remaining fraction of DM. For this, we first determined the total stellar mass of the galaxy (including tidal tails). The colors were obtained by integrating the light of galaxy and tidal tails for each of the bands, $g$, $r$ and $i$. We used the relations between color and mass-to-light ratios from \citet{Roediger2015} for three combinations of colors $g-r$, $g-i$ and $r-i$. The total stellar mass of the galaxy is $M_*=3.6\pm1.3\times10^{7}M_{\odot}$, the average and dispersion from the three individual estimates, at 13.5 Mpc. 

Using the stellar mass to halo mass relationship in \citet{Brook2014}, we derived an expected initial total mass for \dff{} of $M_{h,ini} = 2.2\pm0.3\times10^{10}M_{\odot}$\footnote{Alternatively, if we use the number of GCs to infer the initial halo mass of this galaxy as in \citet{Saifollahi2020}, the result is $M_{h,ini} = 3.6\times10^{10} M_{\odot}$ for the 7 confirmed GCs in \citetalias{vD2019} or $M_{h,ini} = 5.6\times10^{10} M_{\odot}$ for all the GCs identified in this work, a factor $1.8-2.5$ higher.}. Taking the current estimate of the total dynamical mass measured in \citetalias{vD2019}, the average remaining fraction of DM of $\lesssim 1\%$. However, the dynamical mass-to-light ratios of isolated UDGs vary widely \citep[e.g.,][]{Toloba2018, Muller2020} affecting our estimate of the initial total mass of the galaxy. So, the remaining DM fraction could be as high as $10\%$. 

This is consistent with the simulations in \citet{Smith2013, Smith2016}. The bulk of stars in a satellite are located in the very central regions of its DM halo. Hence, the stars are more \emph{shielded} from the tidal forces of the host galaxy with respect to the outer parts of the DM halo. This means that, during the interaction, the DM from the outer parts will be preferentially stripped over the stars of the galaxy \citep{Smith2016}. In other words, the mass stripping of DM haloes proceeds gradually from the outside in \citep[e.g.,][]{Stoehr2002, Penarrubia2008, Smith2016}. 

In \dff, the fraction of stellar mass in the tidal tails is $\sim7\%$ (Sec. \ref{sec:mass_tails}), meaning that the bound stellar mass is $93\%$, in contrast with the $\lesssim1\%$ of DM still bound to the galaxy. This indicates that the galaxy has lost significant amounts of DM before it started losing its stars, in agreement with simulations\footnote{Even if the fraction of stellar mass unbound was of $50\%$ of the current stellar mass of the galaxy, this scenario will still be true.}. Therefore, tidal stripping is a very likely explanation for the extremely low DM content of \dff{} \citep[see also][]{Ogiya2018, Nusser2020, Yang2020, Jackson2020, Maccio2020}. 

Note that we are assuming that the measured DM content of this galaxy in \citetalias{vD2019} is not affected by the tidal interaction. According to \citet{Smith2013}, the observed GC velocity dispersion can be used to measure the true enclosed total mass to within a factor of 2, even when the remaining DM fraction falls as low as $\sim3\%$. However, when it drops below $3\%$, the total mass is overestimated due to the presence of large numbers of unbound GCs that boost the inferred velocity dispersion of the system \citep{Smith2013}. 

To sum up, the preferential tidal stripping of DM over the stars in \dff{} is a very plausible explanation for the measured low content of DM in this galaxy.

\section{Conclusions}
The recent discovery of a second galaxy ``missing'' dark matter had the potential to revolutionized how we understand galaxy formation. In this work, we have shown evidence that this galaxy is undergoing tidal disruption. 
Therefore, \emph{the ``absence of dark matter" in \dff{} is almost certainly caused by its interaction with its neighbour}, NGC1035. Dark matter is less concentrated than stars, and therefore during interactions is preferentially stripped from satellites galaxies. In the case of \dff{}, as the central parts of the galaxy remain untouched and only $\sim7\%$ of the stellar mass of the galaxy is in the tidal tails, we can assume that the stellar component is now starting to be stripped.

\vspace{10mm}

This work shows the necessity of very deep imaging in order to detect and characterize these faint substructures around dwarf satellites in nearby galaxies in order to explain peculiarities that, otherwise, in shallower data remain inexplicable.

\acknowledgments
We would like to thank the anonymous referee for their comments that helped to improve the manuscript. MM thanks Sarah Brough for many useful discussions while preparing this manuscript. We acknowledge support from grant PID2019-107427GB-C32 from The Spanish Ministry of Science and Innovation. We acknowledge financial support from the European Union's Horizon 2020 research and innovation programme under Marie Sk\l odowska-Curie grant agreement No 721463 to the SUNDIAL ITN network, and the European Regional Development Fund (FEDER), from IAC project P/300624, financed by the Ministry of Science, Innovation and Universities, through the State Budget and by the Canary Islands Department of Economy, Knowledge and Employment, through the Regional Budget of the Autonomous Community, and from the Fundaci\'on BBVA under its 2017 programme of assistance to scientific research groups, for the project ``Using machine-learning techniques to drag galaxies from the noise in deep imaging''.
J. R. acknowledge financial support from the grants AYA2015-65973-C3-1-R and RTI2018-096228- B-C31 (MINECO/FEDER, UE), as well as from the State Agency for Research of the Spanish MCIU through the “Center of Excellence Severo Ochoa” award to the Instituto de Astrofísica de Andalucía (SEV-2017-0709).
A. B. was supported by an appointment to the NASA Postdoctoral Program at the NASA Ames Research Center, administered by Universities Space Research Association under contract with NASA.
Based on observations made with the GTC telescope, in the Spanish Observatorio del Roque de los Muchachos of the Instituto de Astrof\'isica de Canarias, under Director’s Discretionary Time.

%

\facilities{HST(ACS), GTC, IAC80}


\software{Astropy \citep{Astropy2018},  
          \sextractor{} \citep{Bertin1996},
          \scamp{} \citep{Bertin2006},
          \swarp{} \citep{Bertin2010},
          Gnuastro \citep{Akhlaghi2015},
          \texttt{photutils} v0.7.2 \citep{Bradley2019},
          \texttt{pillow} \citep{pillow2020},
          \texttt{ellipse} \citep{Jedrzejewski1987},
          numpy \citep{oliphant2006},
          scipy \citep{scipy2020},
          Astrodrizzle \citep{Gonzaga2012},
          Astrometry.net \citep{Lang2010}}

\newpage

\appendix

\section{Surface brightness profiles of the IAC80 images}\label{app:profiles}
\begin{figure*}
   \centering
   \includegraphics[width = \textwidth]{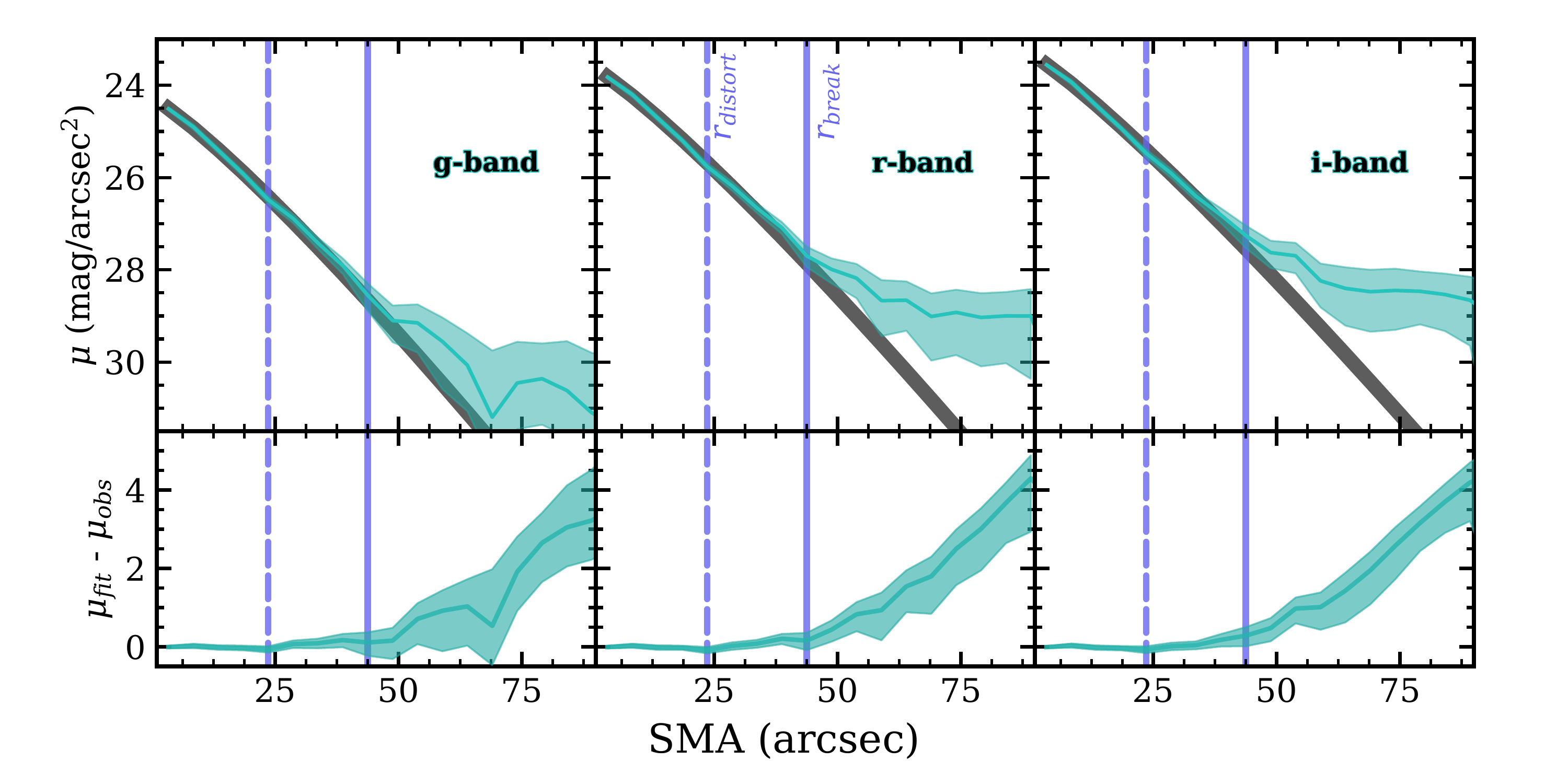}
   \caption{1-D radial surface brightness profiles of the IAC80 images in the $g$ (left), $r$ (middle) and $i$ bands. The lower panels show the residual from fitting the corresponding profile with a S\'ersic model. There is an excess of light beyond $\sim45\arcsec$ in all bands. } \label{fig:profiles}%
    \end{figure*}

Fig. \ref{fig:profiles} show the individual surface brightness profiles derived with \texttt{ellipse} for the IAC80 $g$, $r$ and $i$ bands (top panels). The lower panels show the residuals of fitting a \citet{Sersic1968} model to the profile of the galaxy (grey thick line in the top panels). The details of the fitting procedure are give in Sec. \ref{sec:profs}.

\bibliography{df4}{}

\begin{thebibliography}{}
\expandafter\ifx\csname natexlab\endcsname\relax\def\natexlab#1{#1}\fi
\providecommand{\url}[1]{\href{#1}{#1}}
\providecommand{\dodoi}[1]{doi:~\href{http://doi.org/#1}{\nolinkurl{#1}}}
\providecommand{\doeprint}[1]{\href{http://ascl.net/#1}{\nolinkurl{http://ascl.net/#1}}}
\providecommand{\doarXiv}[1]{\href{https://arxiv.org/abs/#1}{\nolinkurl{https://arxiv.org/abs/#1}}}

\bibitem[{{Aihara} {et~al.}(2018){Aihara}, {Armstrong}, {Bickerton}, {Bosch},
  {Coupon}, {Furusawa}, {Hayashi}, {Ikeda}, {Kamata}, {Karoji}, {Kawanomoto},
  {Koike}, {Komiyama}, {Lang}, {Lupton}, {Mineo}, {Miyatake}, {Miyazaki},
  {Morokuma}, {Obuchi}, {Oishi}, {Okura}, {Price}, {Takata}, {Tanaka},
  {Tanaka}, {Tanaka}, {Uchida}, {Uraguchi}, {Utsumi}, {Wang}, {Yamada},
  {Yamanoi}, {Yasuda}, {Arimoto}, {Chiba}, {Finet}, {Fujimori}, {Fujimoto},
  {Furusawa}, {Goto}, {Goulding}, {Gunn}, {Harikane}, {Hattori}, {Hayashi},
  {He{\l}miniak}, {Higuchi}, {Hikage}, {Ho}, {Hsieh}, {Huang}, {Huang},
  {Imanishi}, {Iwata}, {Jaelani}, {Jian}, {Kashikawa}, {Katayama}, {Kojima},
  {Konno}, {Koshida}, {Kusakabe}, {Leauthaud}, {Lee}, {Lin}, {Lin},
  {Mandelbaum}, {Matsuoka}, {Medezinski}, {Miyama}, {Momose}, {More}, {More},
  {Mukae}, {Murata}, {Murayama}, {Nagao}, {Nakata}, {Niida}, {Niikura},
  {Nishizawa}, {Oguri}, {Okabe}, {Ono}, {Onodera}, {Onoue}, {Ouchi}, {Pyo},
  {Shibuya}, {Shimasaku}, {Simet}, {Speagle}, {Spergel}, {Strauss}, {Sugahara},
  {Sugiyama}, {Suto}, {Suzuki}, {Tait}, {Takada}, {Terai}, {Toba}, {Turner},
  {Uchiyama}, {Umetsu}, {Urata}, {Usuda}, {Yeh}, \& {Yuma}}]{Aihara2018}
{Aihara}, H., {Armstrong}, R., {Bickerton}, S., {et~al.} 2018, \pasj, 70, S8,
  \dodoi{10.1093/pasj/psx081}

\bibitem[{{Akhlaghi}(2019)}]{Akhlaghi2019}
{Akhlaghi}, M. 2019, arXiv e-prints, arXiv:1909.11230.
\newblock \doarXiv{1909.11230}

\bibitem[{{Akhlaghi} \& {Ichikawa}(2015)}]{Akhlaghi2015}
{Akhlaghi}, M., \& {Ichikawa}, T. 2015, \apjs, 220, 1,
  \dodoi{10.1088/0067-0049/220/1/1}

\bibitem[{{Akhlaghi} {et~al.}(2020){Akhlaghi}, {Infante-Sainz}, {Roukema},
  {Valls-Gabaud}, \& {Baena-Gall{\'e}}}]{Akhlaghi2020}
{Akhlaghi}, M., {Infante-Sainz}, R., {Roukema}, B.~F., {Valls-Gabaud}, D., \&
  {Baena-Gall{\'e}}, R. 2020, arXiv e-prints, arXiv:2006.03018.
\newblock \doarXiv{2006.03018}

\bibitem[{{Alam} {et~al.}(2015){Alam}, {Albareti}, {Allende Prieto}, {Anders},
  {Anderson}, {Anderton}, {Andrews}, {Armengaud}, {Aubourg}, {Bailey}, {Basu},
  {Bautista}, {Beaton}, {Beers}, {Bender}, {Berlind}, {Beutler}, {Bhardwaj},
  {Bird}, {Bizyaev}, {Blake}, {Blanton}, {Blomqvist}, {Bochanski}, {Bolton},
  {Bovy}, {Shelden Bradley}, {Brandt}, {Brauer}, {Brinkmann}, {Brown},
  {Brownstein}, {Burden}, {Burtin}, {Busca}, {Cai}, {Capozzi}, {Carnero
  Rosell}, {Carr}, {Carrera}, {Chambers}, {Chaplin}, {Chen}, {Chiappini},
  {Chojnowski}, {Chuang}, {Clerc}, {Comparat}, {Covey}, {Croft}, {Cuesta},
  {Cunha}, {da Costa}, {Da Rio}, {Davenport}, {Dawson}, {De Lee}, {Delubac},
  {Deshpande}, {Dhital}, {Dutra-Ferreira}, {Dwelly}, {Ealet}, {Ebelke},
  {Edmondson}, {Eisenstein}, {Ellsworth}, {Elsworth}, {Epstein}, {Eracleous},
  {Escoffier}, {Esposito}, {Evans}, {Fan}, {Fern{\'a}ndez-Alvar}, {Feuillet},
  {Filiz Ak}, {Finley}, {Finoguenov}, {Flaherty}, {Fleming}, {Font-Ribera},
  {Foster}, {Frinchaboy}, {Galbraith-Frew}, {Garc{\'\i}a},
  {Garc{\'\i}a-Hern{\'a}ndez}, {Garc{\'\i}a P{\'e}rez}, {Gaulme}, {Ge},
  {G{\'e}nova-Santos}, {Georgakakis}, {Ghezzi}, {Gillespie}, {Girardi},
  {Goddard}, {Gontcho}, {Gonz{\'a}lez Hern{\'a}ndez}, {Grebel}, {Green},
  {Grieb}, {Grieves}, {Gunn}, {Guo}, {Harding}, {Hasselquist}, {Hawley},
  {Hayden}, {Hearty}, {Hekker}, {Ho}, {Hogg}, {Holley-Bockelmann}, {Holtzman},
  {Honscheid}, {Huber}, {Huehnerhoff}, {Ivans}, {Jiang}, {Johnson},
  {Kinemuchi}, {Kirkby}, {Kitaura}, {Klaene}, {Knapp}, {Kneib}, {Koenig},
  {Lam}, {Lan}, {Lang}, {Laurent}, {Le Goff}, {Leauthaud}, {Lee}, {Lee},
  {Licquia}, {Liu}, {Long}, {L{\'o}pez-Corredoira}, {Lorenzo-Oliveira},
  {Lucatello}, {Lundgren}, {Lupton}, {Mack}, {Mahadevan}, {Maia}, {Majewski},
  {Malanushenko}, {Malanushenko}, {Manchado}, {Manera}, {Mao}, {Maraston},
  {Marchwinski}, {Margala}, {Martell}, {Martig}, {Masters}, {Mathur},
  {McBride}, {McGehee}, {McGreer}, {McMahon}, {M{\'e}nard}, {Menzel},
  {Merloni}, {M{\'e}sz{\'a}ros}, {Miller}, {Miralda-Escud{\'e}}, {Miyatake},
  {Montero-Dorta}, {More}, {Morganson}, {Morice-Atkinson}, {Morrison},
  {Mosser}, {Muna}, {Myers}, {Nand ra}, {Newman}, {Neyrinck}, {Nguyen},
  {Nichol}, {Nidever}, {Noterdaeme}, {Nuza}, {O'Connell}, {O'Connell},
  {O'Connell}, {Ogando}, {Olmstead}, {Oravetz}, {Oravetz}, {Osumi}, {Owen},
  {Padgett}, {Padmanabhan}, {Paegert}, {Palanque-Delabrouille}, {Pan},
  {Parejko}, {P{\^a}ris}, {Park}, {Pattarakijwanich}, {Pellejero-Ibanez},
  {Pepper}, {Percival}, {P{\'e}rez-Fournon}, {Ṕrez-Ra`fols}, {Petitjean},
  {Pieri}, {Pinsonneault}, {Porto de Mello}, {Prada}, {Prakash},
  {Price-Whelan}, {Protopapas}, {Raddick}, {Rahman}, {Reid}, {Rich}, {Rix},
  {Robin}, {Rockosi}, {Rodrigues}, {Rodr{\'\i}guez-Torres}, {Roe}, {Ross},
  {Ross}, {Rossi}, {Ruan}, {Rubi{\~n}o-Mart{\'\i}n}, {Rykoff},
  {Salazar-Albornoz}, {Salvato}, {Samushia}, {S{\'a}nchez}, {Santiago},
  {Sayres}, {Schiavon}, {Schlegel}, {Schmidt}, {Schneider}, {Schultheis},
  {Schwope}, {Sc{\'o}ccola}, {Scott}, {Sellgren}, {Seo}, {Serenelli}, {Shane},
  {Shen}, {Shetrone}, {Shu}, {Silva Aguirre}, {Sivarani}, {Skrutskie},
  {Slosar}, {Smith}, {Sobreira}, {Souto}, {Stassun}, {Steinmetz}, {Stello},
  {Strauss}, {Streblyanska}, {Suzuki}, {Swanson}, {Tan}, {Tayar}, {Terrien},
  {Thakar}, {Thomas}, {Thomas}, {Thompson}, {Tinker}, {Tojeiro}, {Troup},
  {Vargas-Maga{\~n}a}, {Vazquez}, {Verde}, {Viel}, {Vogt}, {Wake}, {Wang},
  {Weaver}, {Weinberg}, {Weiner}, {White}, {Wilson}, {Wisniewski},
  {Wood-Vasey}, {Ye`che}, {York}, {Zakamska}, {Zamora}, {Zasowski}, {Zehavi},
  {Zhao}, {Zheng}, {Zhou}, {Zhou}, {Zou}, \& {Zhu}}]{Lam2015}
{Alam}, S., {Albareti}, F.~D., {Allende Prieto}, C., {et~al.} 2015, \apjs, 219,
  12, \dodoi{10.1088/0067-0049/219/1/12}

\bibitem[{{Bertin}(2006)}]{Bertin2006}
{Bertin}, E. 2006, in Astronomical Society of the Pacific Conference Series,
  Vol. 351, Astronomical Data Analysis Software and Systems XV, ed.
  C.~{Gabriel}, C.~{Arviset}, D.~{Ponz}, \& S.~{Enrique}, 112

\bibitem[{{Bertin}(2010)}]{Bertin2010}
{Bertin}, E. 2010, {SWarp: Resampling and Co-adding FITS Images Together}.
\newblock \doeprint{1010.068}

\bibitem[{{Bertin} \& {Arnouts}(1996)}]{Bertin1996}
{Bertin}, E., \& {Arnouts}, S. 1996, \aaps, 117, 393,
  \dodoi{10.1051/aas:1996164}

\bibitem[{{Borlaff} {et~al.}(2019){Borlaff}, {Trujillo}, {Rom{\'a}n},
  {Beckman}, {Eliche-Moral}, {Infante-S{\'a}inz}, {Lumbreras-Calle}, {de
  Almagro}, {G{\'o}mez-Guijarro}, {Cebri{\'a}n}, {Dorta}, {Cardiel},
  {Akhlaghi}, \& {Mart{\'\i}nez-Lombilla}}]{Borlaff2019}
{Borlaff}, A., {Trujillo}, I., {Rom{\'a}n}, J., {et~al.} 2019, \aap, 621, A133,
  \dodoi{10.1051/0004-6361/201834312}

\bibitem[{Bradley {et~al.}(2019)Bradley, Sipocz, Robitaille, Tollerud,
  Vinícius, Deil, Barbary, Busko, Günther, Cara, Wilson, Conseil, Droettboom,
  Bostroem, Bray, Bratholm, Lim, Craig, Barentsen, Pascual, Donath, Greco,
  Perren, Kerzendorf, de~Val-Borro, Dencheva, de~Albernaz~Ferreira, Souchereau,
  D'Eugenio, \& Weaver}]{Bradley2019}
Bradley, L., Sipocz, B., Robitaille, T., {et~al.} 2019, astropy/photutils:
  v0.7.1, v0.7.1,  Zenodo, \dodoi{10.5281/zenodo.3478575}

\bibitem[{{Brodie} \& {Strader}(2006)}]{Brodie2006}
{Brodie}, J.~P., \& {Strader}, J. 2006, \araa, 44, 193,
  \dodoi{10.1146/annurev.astro.44.051905.092441}

\bibitem[{{Brook} {et~al.}(2014){Brook}, {Di Cintio}, {Knebe}, {Gottl{\"o}ber},
  {Hoffman}, {Yepes}, \& {Garrison-Kimmel}}]{Brook2014}
{Brook}, C.~B., {Di Cintio}, A., {Knebe}, A., {et~al.} 2014, \apjl, 784, L14,
  \dodoi{10.1088/2041-8205/784/1/L14}

\bibitem[{{Cohen} {et~al.}(2018){Cohen}, {van Dokkum}, {Danieli}, {Romanowsky},
  {Abraham}, {Merritt}, {Zhang}, {Mowla}, {Kruijssen}, {Conroy}, \&
  {Wasserman}}]{Cohen2018}
{Cohen}, Y., {van Dokkum}, P., {Danieli}, S., {et~al.} 2018, \apj, 868, 96,
  \dodoi{10.3847/1538-4357/aae7c8}

\bibitem[{{Danieli} {et~al.}(2020){Danieli}, {van Dokkum}, {Abraham}, {Conroy},
  {Dolphin}, \& {Romanowsky}}]{Danieli2020}
{Danieli}, S., {van Dokkum}, P., {Abraham}, R., {et~al.} 2020, \apjl, 895, L4,
  \dodoi{10.3847/2041-8213/ab8dc4}

\bibitem[{{Dhillon} {et~al.}(2018){Dhillon}, {Dixon}, {Gamble}, {Kerry},
  {Littlefair}, {Parsons}, {Marsh}, {Bezawada}, {Black}, {Gao}, {Henry},
  {Lunney}, {Miller}, {Dubbeldam}, {Morris}, {Osborn}, {Wilson}, {Casares},
  {Mu{\~n}oz-Darias}, {Pall{\'e}}, {Rodriguez-Gil}, {Shahbaz}, \& {de Ugarte
  Postigo}}]{Dhillon2018}
{Dhillon}, V., {Dixon}, S., {Gamble}, T., {et~al.} 2018, in Society of
  Photo-Optical Instrumentation Engineers (SPIE) Conference Series, Vol. 10702,
  \procspie, 107020L, \dodoi{10.1117/12.2312041}

\bibitem[{{Duc} {et~al.}(2000){Duc}, {Brinks}, {Springel}, {Pichardo},
  {Weilbacher}, \& {Mirabel}}]{Duc2000}
{Duc}, P.~A., {Brinks}, E., {Springel}, V., {et~al.} 2000, \aj, 120, 1238,
  \dodoi{10.1086/301516}

\bibitem[{{Forbes}(2017)}]{Forbes2017}
{Forbes}, D.~A. 2017, \mnras, 472, L104, \dodoi{10.1093/mnrasl/slx148}

\bibitem[{{Freeman}(1970)}]{Freeman1970}
{Freeman}, K.~C. 1970, \apj, 160, 811, \dodoi{10.1086/150474}

\bibitem[{{Gonzaga} {et~al.}(2012){Gonzaga}, {Hack}, {Fruchter}, \&
  {Mack}}]{Gonzaga2012}
{Gonzaga}, S., {Hack}, W., {Fruchter}, A., \& {Mack}, J. 2012, {The DrizzlePac
  Handbook}

\bibitem[{{Goranova} {et~al.}(2009){Goranova}, {Hudelot}, {Magnard}, \&
  et~al.}]{Goranova2009}
{Goranova}, Y., {Hudelot}, P., {Magnard}, F., \& et~al. 2009, {The CFHTLS T0006
  Release}.
\newblock \url{http://terapix.iap.fr/cplt/T0006-doc.pdf}

\bibitem[{{Harris}(1996)}]{Harris1996}
{Harris}, W.~E. 1996, \aj, 112, 1487, \dodoi{10.1086/118116}

\bibitem[{{Hudson} \& {Robison}(2018)}]{Hudson2018}
{Hudson}, M.~J., \& {Robison}, B. 2018, \mnras, 477, 3869,
  \dodoi{10.1093/mnras/sty844}

\bibitem[{{Hughes} {et~al.}(2019){Hughes}, {Pfeffer}, {Martig}, {Bastian},
  {Crain}, {Kruijssen}, \& {Reina-Campos}}]{Hughes2019}
{Hughes}, M.~E., {Pfeffer}, J., {Martig}, M., {et~al.} 2019, \mnras, 482, 2795,
  \dodoi{10.1093/mnras/sty2889}

\bibitem[{{Infante-Sainz} {et~al.}(2020){Infante-Sainz}, {Trujillo}, \&
  {Rom{\'a}n}}]{Infante2020}
{Infante-Sainz}, R., {Trujillo}, I., \& {Rom{\'a}n}, J. 2020, \mnras, 491,
  5317, \dodoi{10.1093/mnras/stz3111}

\bibitem[{{Jackson} {et~al.}(2020){Jackson}, {Kaviraj}, {Martin}, {Devriendt},
  {Slyz}, {Silk}, {Dubois}, {Yi}, {Pichon}, {Volonteri}, {Choi}, {Kimm},
  {Kraljic}, \& {Peirani}}]{Jackson2020}
{Jackson}, R.~A., {Kaviraj}, S., {Martin}, G., {et~al.} 2020, arXiv e-prints,
  arXiv:2010.02219.
\newblock \doarXiv{2010.02219}

\bibitem[{{Jedrzejewski}(1987)}]{Jedrzejewski1987}
{Jedrzejewski}, R.~I. 1987, \mnras, 226, 747, \dodoi{10.1093/mnras/226.4.747}

\bibitem[{{Johnston} {et~al.}(2002){Johnston}, {Choi}, \&
  {Guhathakurta}}]{Johnston2002}
{Johnston}, K.~V., {Choi}, P.~I., \& {Guhathakurta}, P. 2002, \aj, 124, 127,
  \dodoi{10.1086/341040}

\bibitem[{{Johnston} {et~al.}(1996){Johnston}, {Hernquist}, \&
  {Bolte}}]{Johnston1996}
{Johnston}, K.~V., {Hernquist}, L., \& {Bolte}, M. 1996, \apj, 465, 278,
  \dodoi{10.1086/177418}

\bibitem[{{Klimentowski} {et~al.}(2009){Klimentowski}, {{\L}okas},
  {Kazantzidis}, {Mayer}, {Mamon}, \& {Prada}}]{Klimentowski2009}
{Klimentowski}, J., {{\L}okas}, E.~L., {Kazantzidis}, S., {et~al.} 2009,
  \mnras, 400, 2162, \dodoi{10.1111/j.1365-2966.2009.15626.x}

\bibitem[{{Koch} {et~al.}(2012){Koch}, {Burkert}, {Rich}, {Collins}, {Black},
  {Hilker}, \& {Benson}}]{Koch2012}
{Koch}, A., {Burkert}, A., {Rich}, R.~M., {et~al.} 2012, \apjl, 755, L13,
  \dodoi{10.1088/2041-8205/755/1/L13}

\bibitem[{{Lang} {et~al.}(2010){Lang}, {Hogg}, {Mierle}, {Blanton}, \&
  {Roweis}}]{Lang2010}
{Lang}, D., {Hogg}, D.~W., {Mierle}, K., {Blanton}, M., \& {Roweis}, S. 2010,
  \aj, 139, 1782, \dodoi{10.1088/0004-6256/139/5/1782}

\bibitem[{{Lelli} {et~al.}(2015){Lelli}, {Duc}, {Brinks}, {Bournaud},
  {McGaugh}, {Lisenfeld}, {Weilbacher}, {Boquien}, {Revaz}, {Braine},
  {Koribalski}, \& {Belles}}]{Lelli2015}
{Lelli}, F., {Duc}, P.-A., {Brinks}, E., {et~al.} 2015, \aap, 584, A113,
  \dodoi{10.1051/0004-6361/201526613}

\bibitem[{{Macci{\`o}} {et~al.}(2020){Macci{\`o}}, {Huterer Prats}, {Dixon},
  {Buck}, {Waterval}, {Arora}, {Courteau}, \& {Kang}}]{Maccio2020}
{Macci{\`o}}, A.~V., {Huterer Prats}, D., {Dixon}, K.~L., {et~al.} 2020, arXiv
  e-prints, arXiv:2010.02245.
\newblock \doarXiv{2010.02245}

\bibitem[{{Mackey} {et~al.}(2010){Mackey}, {Huxor}, {Ferguson}, {Irwin},
  {Tanvir}, {McConnachie}, {Ibata}, {Chapman}, \& {Lewis}}]{Mackey2010}
{Mackey}, A.~D., {Huxor}, A.~P., {Ferguson}, A.~M.~N., {et~al.} 2010, \apjl,
  717, L11, \dodoi{10.1088/2041-8205/717/1/L11}

\bibitem[{{Merritt} {et~al.}(2016){Merritt}, {van Dokkum}, {Abraham}, \&
  {Zhang}}]{Merritt2016}
{Merritt}, A., {van Dokkum}, P., {Abraham}, R., \& {Zhang}, J. 2016, \apj, 830,
  62, \dodoi{10.3847/0004-637X/830/2/62}

\bibitem[{{Monelli} \& {Trujillo}(2019)}]{Monelli2019}
{Monelli}, M., \& {Trujillo}, I. 2019, \apjl, 880, L11,
  \dodoi{10.3847/2041-8213/ab2fd2}

\bibitem[{{Montes} {et~al.}(2014){Montes}, {Acosta-Pulido}, {Prieto}, \&
  {Fern{\'a}ndez-Ontiveros}}]{Montes2014}
{Montes}, M., {Acosta-Pulido}, J.~A., {Prieto}, M.~A., \&
  {Fern{\'a}ndez-Ontiveros}, J.~A. 2014, \mnras, 442, 1350,
  \dodoi{10.1093/mnras/stu948}

\bibitem[{{Montes} \& {Trujillo}(2014)}]{MT14}
{Montes}, M., \& {Trujillo}, I. 2014, \apj, 794, 137,
  \dodoi{10.1088/0004-637X/794/2/137}

\bibitem[{{Montes} \& {Trujillo}(2019)}]{MT19}
---. 2019, \mnras, 482, 2838, \dodoi{10.1093/mnras/sty2858}

\bibitem[{{Mu{\~n}oz} {et~al.}(2014){Mu{\~n}oz}, {Puzia}, {Lan{\c{c}}on},
  {Peng}, {C{\^o}t{\'e}}, {Ferrarese}, {Blakeslee}, {Mei}, {Cuillandre},
  {Hudelot}, {Courteau}, {Duc}, {Balogh}, {Boselli}, {Bournaud}, {Carlberg},
  {Chapman}, {Durrell}, {Eigenthaler}, {Emsellem}, {Gavazzi}, {Gwyn},
  {Huertas-Company}, {Ilbert}, {Jord{\'a}n}, {L{\"a}sker}, {Licitra}, {Liu},
  {MacArthur}, {McConnachie}, {McCracken}, {Mellier}, {Peng}, {Raichoor},
  {Taylor}, {Tonry}, {Tully}, \& {Zhang}}]{Munoz2014}
{Mu{\~n}oz}, R.~P., {Puzia}, T.~H., {Lan{\c{c}}on}, A., {et~al.} 2014, \apjs,
  210, 4, \dodoi{10.1088/0067-0049/210/1/4}

\bibitem[{{M{\"u}ller} {et~al.}(2019){M{\"u}ller}, {Rich}, {Rom{\'a}n},
  {Y{\i}ld{\i}z}, {B{\'\i}lek}, {Duc}, {Fensch}, {Trujillo}, \&
  {Koch}}]{Muller2019}
{M{\"u}ller}, O., {Rich}, R.~M., {Rom{\'a}n}, J., {et~al.} 2019, \aap, 624, L6,
  \dodoi{10.1051/0004-6361/201935463}

\bibitem[{{M{\"u}ller} {et~al.}(2020){M{\"u}ller}, {Marleau}, {Duc}, {Habas},
  {Fensch}, {Emsellem}, {Poulain}, {Lim}, {Agnello}, {Durrell}, {Paudel},
  {S{\'a}nchez-Janssen}, \& {van der Burg}}]{Muller2020}
{M{\"u}ller}, O., {Marleau}, F.~R., {Duc}, P.-A., {et~al.} 2020, \aap, 640,
  A106, \dodoi{10.1051/0004-6361/202038351}

\bibitem[{{Nusser}(2020)}]{Nusser2020}
{Nusser}, A. 2020, \apj, 893, 66, \dodoi{10.3847/1538-4357/ab792c}

\bibitem[{{Ogiya}(2018)}]{Ogiya2018}
{Ogiya}, G. 2018, \mnras, 480, L106, \dodoi{10.1093/mnrasl/sly138}

\bibitem[{Oliphant(2006)}]{oliphant2006}
Oliphant, T.~E. 2006, A guide to NumPy, Vol.~1 (Trelgol Publishing USA)

\bibitem[{{Pe{\~n}arrubia} {et~al.}(2008){Pe{\~n}arrubia}, {Navarro}, \&
  {McConnachie}}]{Penarrubia2008}
{Pe{\~n}arrubia}, J., {Navarro}, J.~F., \& {McConnachie}, A.~W. 2008, \apj,
  673, 226, \dodoi{10.1086/523686}

\bibitem[{{Pe{\~n}arrubia} {et~al.}(2009){Pe{\~n}arrubia}, {Navarro},
  {McConnachie}, \& {Martin}}]{Penarrubia2009}
{Pe{\~n}arrubia}, J., {Navarro}, J.~F., {McConnachie}, A.~W., \& {Martin},
  N.~F. 2009, \apj, 698, 222, \dodoi{10.1088/0004-637X/698/1/222}

\bibitem[{{Pierce} {et~al.}(2005){Pierce}, {Brodie}, {Forbes}, {Beasley},
  {Proctor}, \& {Strader}}]{Pierce2005}
{Pierce}, M., {Brodie}, J.~P., {Forbes}, D.~A., {et~al.} 2005, \mnras, 358,
  419, \dodoi{10.1111/j.1365-2966.2005.08778.x}

\bibitem[{{Ploeckinger} {et~al.}(2015){Ploeckinger}, {Recchi}, {Hensler}, \&
  {Kroupa}}]{Ploeckinger2015}
{Ploeckinger}, S., {Recchi}, S., {Hensler}, G., \& {Kroupa}, P. 2015, \mnras,
  447, 2512, \dodoi{10.1093/mnras/stu2629}

\bibitem[{{Ploeckinger} {et~al.}(2018){Ploeckinger}, {Sharma}, {Schaye},
  {Crain}, {Schaller}, \& {Barber}}]{Ploeckinger2018}
{Ploeckinger}, S., {Sharma}, K., {Schaye}, J., {et~al.} 2018, \mnras, 474, 580,
  \dodoi{10.1093/mnras/stx2787}

\bibitem[{{Read} {et~al.}(2006){Read}, {Wilkinson}, {Evans}, {Gilmore}, \&
  {Kleyna}}]{Read2006}
{Read}, J.~I., {Wilkinson}, M.~I., {Evans}, N.~W., {Gilmore}, G., \& {Kleyna},
  J.~T. 2006, \mnras, 367, 387, \dodoi{10.1111/j.1365-2966.2005.09959.x}

\bibitem[{{Rix} {et~al.}(2004){Rix}, {Barden}, {Beckwith}, {Bell}, {Borch},
  {Caldwell}, {H{\"a}ussler}, {Jahnke}, {Jogee}, {McIntosh}, {Meisenheimer},
  {Peng}, {Sanchez}, {Somerville}, {Wisotzki}, \& {Wolf}}]{Rix2004}
{Rix}, H.-W., {Barden}, M., {Beckwith}, S. V.~W., {et~al.} 2004, \apjs, 152,
  163, \dodoi{10.1086/420885}

\bibitem[{{Roediger} \& {Courteau}(2015)}]{Roediger2015}
{Roediger}, J.~C., \& {Courteau}, S. 2015, \mnras, 452, 3209,
  \dodoi{10.1093/mnras/stv1499}

\bibitem[{{Rom{\'a}n} {et~al.}(2019){Rom{\'a}n}, {Trujillo}, \&
  {Montes}}]{Roman2019}
{Rom{\'a}n}, J., {Trujillo}, I., \& {Montes}, M. 2019, arXiv e-prints,
  arXiv:1907.00978.
\newblock \doarXiv{1907.00978}

\bibitem[{{Rubin} \& {Ford}(1970)}]{Rubin1970}
{Rubin}, V.~C., \& {Ford}, W.~Kent, J. 1970, \apj, 159, 379,
  \dodoi{10.1086/150317}

\bibitem[{{Saifollahi} {et~al.}(2020){Saifollahi}, {Trujillo}, {Beasley},
  {Peletier}, \& {Knapen}}]{Saifollahi2020}
{Saifollahi}, T., {Trujillo}, I., {Beasley}, M.~A., {Peletier}, R.~F., \&
  {Knapen}, J.~H. 2020, arXiv e-prints, arXiv:2006.14630.
\newblock \doarXiv{2006.14630}

\bibitem[{{S\'{e}rsic}(1968)}]{Sersic1968}
{S\'{e}rsic}, J.~L. 1968, {Atlas de galaxias australes}

\bibitem[{{Shen} {et~al.}(2020){Shen}, {van Dokkum}, \& {Danieli}}]{Shen2020}
{Shen}, Z., {van Dokkum}, P., \& {Danieli}, S. 2020, arXiv e-prints,
  arXiv:2010.07324.
\newblock \doarXiv{2010.07324}

\bibitem[{{Smith} {et~al.}(2016){Smith}, {Choi}, {Lee}, {Rhee},
  {Sanchez-Janssen}, \& {Yi}}]{Smith2016}
{Smith}, R., {Choi}, H., {Lee}, J., {et~al.} 2016, \apj, 833, 109,
  \dodoi{10.3847/1538-4357/833/1/109}

\bibitem[{{Smith} {et~al.}(2013){Smith}, {S{\'a}nchez-Janssen}, {Fellhauer},
  {Puzia}, {Aguerri}, \& {Farias}}]{Smith2013}
{Smith}, R., {S{\'a}nchez-Janssen}, R., {Fellhauer}, M., {et~al.} 2013, \mnras,
  429, 1066, \dodoi{10.1093/mnras/sts395}

\bibitem[{{Sorce} {et~al.}(2014){Sorce}, {Tully}, {Courtois}, {Jarrett},
  {Neill}, \& {Shaya}}]{Sorce2014}
{Sorce}, J.~G., {Tully}, R.~B., {Courtois}, H.~M., {et~al.} 2014, \mnras, 444,
  527, \dodoi{10.1093/mnras/stu1450}

\bibitem[{{Stoehr} {et~al.}(2002){Stoehr}, {White}, {Tormen}, \&
  {Springel}}]{Stoehr2002}
{Stoehr}, F., {White}, S. D.~M., {Tormen}, G., \& {Springel}, V. 2002, \mnras,
  335, L84, \dodoi{10.1046/j.1365-8711.2002.05891.x}

\bibitem[{{Taylor} {et~al.}(2017){Taylor}, {Puzia}, {Mu{\~n}oz}, {Mieske},
  {Lan{\c{c}}on}, {Zhang}, {Eigenthaler}, \& {Bovill}}]{Taylor2017}
{Taylor}, M.~A., {Puzia}, T.~H., {Mu{\~n}oz}, R.~P., {et~al.} 2017, \mnras,
  469, 3444, \dodoi{10.1093/mnras/stx1021}

\bibitem[{{The Astropy Collaboration} {et~al.}(2018){The Astropy
  Collaboration}, {Price-Whelan}, {Sip{\H o}cz}, {G{\"u}nther}, {Lim},
  {Crawford}, {Conseil}, {Shupe}, {Craig}, {Dencheva}, {Ginsburg},
  {VanderPlas}, {Bradley}, {P{\'e}rez-Su{\'a}rez}, {de Val-Borro}, {Aldcroft},
  {Cruz}, {Robitaille}, {Tollerud}, {Ardelean}, {Babej}, {Bachetti}, {Bakanov},
  {Bamford}, {Barentsen}, {Barmby}, {Baumbach}, {Berry}, {Biscani}, {Boquien},
  {Bostroem}, {Bouma}, {Brammer}, {Bray}, {Breytenbach}, {Buddelmeijer},
  {Burke}, {Calderone}, {Cano Rodr{\'{\i}}guez}, {Cara}, {Cardoso},
  {Cheedella}, {Copin}, {Crichton}, {D{\'A}vella}, {Deil}, {Depagne},
  {Dietrich}, {Donath}, {Droettboom}, {Earl}, {Erben}, {Fabbro}, {Ferreira},
  {Finethy}, {Fox}, {Garrison}, {Gibbons}, {Goldstein}, {Gommers}, {Greco},
  {Greenfield}, {Groener}, {Grollier}, {Hagen}, {Hirst}, {Homeier}, {Horton},
  {Hosseinzadeh}, {Hu}, {Hunkeler}, {Ivezi{\'c}}, {Jain}, {Jenness}, {Kanarek},
  {Kendrew}, {Kern}, {Kerzendorf}, {Khvalko}, {King}, {Kirkby}, {Kulkarni},
  {Kumar}, {Lee}, {Lenz}, {Littlefair}, {Ma}, {Macleod}, {Mastropietro},
  {McCully}, {Montagnac}, {Morris}, {Mueller}, {Mumford}, {Muna}, {Murphy},
  {Nelson}, {Nguyen}, {Ninan}, {N{\"o}the}, {Ogaz}, {Oh}, {Parejko}, {Parley},
  {Pascual}, {Patil}, {Patil}, {Plunkett}, {Prochaska}, {Rastogi}, {Reddy
  Janga}, {Sabater}, {Sakurikar}, {Seifert}, {Sherbert}, {Sherwood-Taylor},
  {Shih}, {Sick}, {Silbiger}, {Singanamalla}, {Singer}, {Sladen}, {Sooley},
  {Sornarajah}, {Streicher}, {Teuben}, {Thomas}, {Tremblay}, {Turner},
  {Terr{\'o}n}, {van Kerkwijk}, {de la Vega}, {Watkins}, {Weaver}, {Whitmore},
  {Woillez}, \& {Zabalza}}]{Astropy2018}
{The Astropy Collaboration}, {Price-Whelan}, A.~M., {Sip{\H o}cz}, B.~M.,
  {et~al.} 2018, ArXiv e-prints.
\newblock \doarXiv{1801.02634}

\bibitem[{{Toloba} {et~al.}(2018){Toloba}, {Lim}, {Peng}, {Sales},
  {Guhathakurta}, {Mihos}, {C{\^o}t{\'e}}, {Boselli}, {Cuillandre},
  {Ferrarese}, {Gwyn}, {Lan{\c{c}}on}, {Mu{\~n}oz}, \& {Puzia}}]{Toloba2018}
{Toloba}, E., {Lim}, S., {Peng}, E., {et~al.} 2018, \apjl, 856, L31,
  \dodoi{10.3847/2041-8213/aab603}

\bibitem[{{Toomre} \& {Toomre}(1972)}]{Toomre1972}
{Toomre}, A., \& {Toomre}, J. 1972, \apj, 178, 623, \dodoi{10.1086/151823}

\bibitem[{{Trujillo} \& {Fliri}(2016)}]{Trujillo2016}
{Trujillo}, I., \& {Fliri}, J. 2016, \apj, 823, 123,
  \dodoi{10.3847/0004-637X/823/2/123}

\bibitem[{{Trujillo} {et~al.}(2019){Trujillo}, {Beasley}, {Borlaff},
  {Carrasco}, {Di Cintio}, {Filho}, {Monelli}, {Montes}, {Rom{\'a}n},
  {Ruiz-Lara}, {S{\'a}nchez Almeida}, {Valls-Gabaud}, \&
  {Vazdekis}}]{Trujillo2019}
{Trujillo}, I., {Beasley}, M.~A., {Borlaff}, A., {et~al.} 2019, \mnras, 486,
  1192, \dodoi{10.1093/mnras/stz771}

\bibitem[{{Truong} {et~al.}(2017){Truong}, {Newman}, {Simon}, {Blitz}, {Ellis},
  \& {Bolatto}}]{Truong2017}
{Truong}, P.~N., {Newman}, A.~B., {Simon}, J.~D., {et~al.} 2017, \apj, 843, 37,
  \dodoi{10.3847/1538-4357/aa76eb}

\bibitem[{{van Dokkum} {et~al.}(2019){van Dokkum}, {Danieli}, {Abraham},
  {Conroy}, \& {Romanowsky}}]{vD2019}
{van Dokkum}, P., {Danieli}, S., {Abraham}, R., {Conroy}, C., \& {Romanowsky},
  A.~J. 2019, \apjl, 874, L5, \dodoi{10.3847/2041-8213/ab0d92}

\bibitem[{{van Dokkum} {et~al.}(2018){van Dokkum}, {Danieli}, {Cohen},
  {Merritt}, {Romanowsky}, {Abraham}, {Brodie}, {Conroy}, {Lokhorst}, {Mowla},
  {O'Sullivan}, \& {Zhang}}]{vD_df2}
{van Dokkum}, P., {Danieli}, S., {Cohen}, Y., {et~al.} 2018, \nat, 555, 629,
  \dodoi{10.1038/nature25767}

\bibitem[{van Kemenade {et~al.}(2020)van Kemenade, wiredfool, Murray, Clark,
  Karpinsky, Gohlke, Dufresne, nulano, Crowell, Schmidt, Houghton, Kopachev,
  Landey, Mani, vashek, Ware, Jason, Caro, Lahd, Kossouho, Bonfill, Tonnhofer,
  Brown, Lee, (변기호), Al-Saidi, Kurczewski, Korobov, Górny, \&
  Yang}]{pillow2020}
van Kemenade, H., wiredfool, Murray, A., {et~al.} 2020, python-pillow/Pillow
  7.1.1, 7.1.1,  Zenodo, \dodoi{10.5281/zenodo.3738618}

\bibitem[{Virtanen {et~al.}(2020)Virtanen, Gommers, Oliphant, Haberland, Reddy,
  Cournapeau, Burovski, Peterson, Weckesser, Bright, {van der Walt}, Brett,
  Wilson, Millman, Mayorov, Nelson, Jones, Kern, Larson, Carey, Polat, Feng,
  Moore, {VanderPlas}, Laxalde, Perktold, Cimrman, Henriksen, Quintero, Harris,
  Archibald, Ribeiro, Pedregosa, {van Mulbregt}, \& {SciPy 1.0
  Contributors}}]{scipy2020}
Virtanen, P., Gommers, R., Oliphant, T.~E., {et~al.} 2020, Nature Methods, 17,
  261, \dodoi{10.1038/s41592-019-0686-2}

\bibitem[{{Yang} {et~al.}(2020){Yang}, {Yu}, \& {An}}]{Yang2020}
{Yang}, D., {Yu}, H.-B., \& {An}, H. 2020, \prl, 125, 111105,
  \dodoi{10.1103/PhysRevLett.125.111105}

\end{thebibliography}
\bibliographystyle{aasjournal}



\end{document}